\documentclass[12pt]{article}
\usepackage[noblocks]{authblk}

\title{Learning To Fold Proteins Using Energy Landscape Theory}
\author[1,3]{N.P. Schafer}
\author[2,3]{B.L. Kim}
\author[2,3]{W. Zheng}
\author[1,2,3]{P.G. Wolynes}
\affil[1]{Department of Physics, Rice University, Houston, TX 77005, USA}
\affil[2]{Department of Chemistry, Rice University, Houston, TX 77005, USA}
\affil[3]{Center for Theoretical Biological Physics, Rice University, Houston, TX 77005, USA}
\date{\today}

\usepackage[scale=0.75]{geometry}
\usepackage{setspace}
\usepackage{amsmath}
\usepackage{bm}
\usepackage{graphicx}

\begin{document}
\maketitle
\newpage
\begin{abstract}
This review is a tutorial for scientists interested in the problem of protein structure prediction, particularly those interested in using coarse-grained molecular dynamics models that are optimized using lessons learned from the energy landscape theory of protein folding. We also present a review of the results of the AMH/AMC/AMW/AWSEM family of coarse-grained molecular dynamics protein folding models to illustrate the points covered in the first part of the article. Accurate coarse-grained structure prediction models can be used to investigate a wide range of conceptual and mechanistic issues outside of protein structure prediction; specifically, the paper concludes by reviewing how AWSEM has in recent years been able to elucidate questions related to the unusual kinetic behavior of artificially designed proteins, multidomain protein misfolding, and the initial stages of protein aggregation.
\end{abstract}

\begin{center}
{\it The sooner you make your first 5000 mistakes, the sooner you will be able to correct them.} \\ -- Nicolaides~\cite{edwards1997drawing}
\end{center}

\newpage

\tableofcontents

\newpage 

\section{The protein structure prediction problem}
\label{sec:definitions}
That proteins fold to organized structures is an essential biological fact and a remarkable physical phenomenon. The conceptual understanding of this phenomenon has captivated theorists of all stripes. The essential paradoxes of how proteins can fold have been resolved by energy landscape theory~\cite{wolynes1996three,wolynes2012chemical,Service08082008} but for practical persons the ``protein folding problem'' is the problem of learning how to predict protein tertiary structure. The practical motivation for being able reliably to predict protein structure is clear - finding the amino acid sequences of proteins is easy and extraordinarily cheap but, despite enormous effort and tremendous technological advances~\cite{chandonia2006impact}, obtaining full three dimensional structures experimentally remains a challenge and is still comparatively expensive. Structure and function are so closely linked in biology that even crude structures give functional insights and highly accurate structures can help understand issues of specificity important to systems biology and medicine. The problem of predicting structure from sequence is sufficiently interesting and important so that over the last few decades structure prediction has become its own sub-field with a myriad of approaches being developed by many members of a diverse community. Because of the open ended nature of the problem, structure prediction technologies often appear {\it ad hoc} and diverse in method. Energy landscape theory provides a consistent way of navigating the methodology labyrinth. The goal of this paper is to lead both neophyte and expert through the structure prediction problem using energy landscape theory. 

How do we actually define the problem of protein structure prediction? Grossly speaking, there are two main types of protein structure prediction approaches: template-based and template-free. Template-based modeling, also known as homology modeling, relies on there being one or more structures already determined of proteins which are sufficiently similar in sequence to the target sequence so that the structure for the target sequence can be predicted by analogy with those already known. Obtaining the starting templates offers severe constraints on the final predictions. When it can be done, template-based modeling is presently the most reliable way of predicting structure. Although the number of experimentally resolved structures is low compared to the number of experimentally determined sequences, many (or even most) sequences are still good candidates for template-based modeling because sequences, although quite different, often yield very similar structures. Nevertheless, a pair of sequences corresponding to very similar structure with between 25-40\% sequence identity may not be recognizable immediately as being candidates for template modeling and are sometimes said to be in the ``twilight zone''~\cite{bowie1991method}. On the other hand, sequences having greater than 40\% sequence identity to another sequence with an experimentally determined structure are usually good candidates for template-based modeling and are easily identified as such. These observations illustrate the fact that structure evolves more slowly than sequence and that many widely differing sequences correspond with the same overall fold. The robustness of protein structures to mutation arises from the funneled nature of the energy landscape~\cite{taverna2002proteins,pande1995accurate}. In this way, the funneled shape of protein energy landscapes lies at the heart of homology modeling even when landscape theory is not explicitly invoked by the practitioners.

While homology modeling will be briefly discussed, {\it en passant}, this review will primarily focus on template-free modeling. Template-free modeling, also known as {\it de novo} or sometimes {\it ab initio} structure prediction, is performed without making explicit use of experimentally resolved structures of known homologous sequences. To some, the term {\it ab initio} structure prediction also connotes only structure prediction using atomistically detailed models starting from basic molecular physics. Yet many aspects of protein structures, those robust between homologs in fact, are transferable only at a coarse-grained level since molecular evolution itself works only at the amino acid level. It has proven quite sensible to decompose protein structure itself into primary, secondary, tertiary and quaternary structure. The translation of DNA sequences into the primary sequence, or amino acid sequence, of a protein polymer is, apart from the existence of introns, a simple symbol translation problem and is a local problem mapping one form of one dimensional information onto another. The secondary structures of proteins are quite regular and the main varieties of local structure are strikingly few in number, as predicted early on by Pauling and others on the basis of simple physical arguments regarding the satisfaction of backbone hydrogen bonding patterns~\cite{pauling1951structure}. Again a one-dimensional mapping would seem sensible. Nevertheless, the prediction of protein secondary structure from sequence information is not entirely local and appears to be inextricably linked to the prediction of tertiary structure~\cite{meiler2003coupled}. A protein's tertiary structure can be thought of as the three dimensional packing of the secondary structural elements, including helices and sheets. This review will first focus primarily on the problem of the prediction of the tertiary structure of single domains. The prediction of quaternary structure, or the relative arrangement and packing of protein tertiary structures, is a still more important problem which is made easier when the tertiary structures of the components are already available and which does share similar features. This involves the search for protein binding sites and interfaces.

For template-free tertiary structure prediction, there are also several ways of evaluating structure prediction outcomes and tasks, which correspond roughly to different levels of ambition or difficulty. The most forgiving version of the structure prediction problem, but a nonetheless important and well studied first step, is that of ``native state recognition'' or threading. Here, in one approach, the sequence of a protein is threaded over the possible tertiary structures as well as experimentally determined tertiary structures corresponding to unrelated sequences~\cite{fischer1996assigning,koretke1996self,goldstein1992protein,goldstein1993three}. If the lowest energy, or highest score, is obtained for the correct positioning of the sequence along the structure, then the native state is said to have been ``recognized''. A large range of approaches have been found to be successful at this level of structure prediction. Such one-pass recognition does not correspond to a complete solution to the structure prediction problem because these methods are only required to discriminate between a tiny subset of somewhat artificial misfolded configurations and the native state, which is is necessarily already assumed to be in the cataloged list of possibilities. Considerably more difficult is the problem of predicting protein structure by assembling fragments from existing proteins (thereby acknowledging the partially local character of a secondary structure code~\cite{saven1996local}) or by carrying out molecular dynamics on a model with a flexible backbone that could take on even local structures that are unprecedented~\cite{friedrichs1990molecular,davtyan2012awsem,fujitsuka2004optimizing,kenzaki2011cafemol}. In energy landscape terms, the reason why the above described types of structure prediction are progressively more difficult can be understood by looking at Figure~\ref{fig:typesofdecoys}, which will be explained in greater detail in Section~\ref{sec:theory} below.
\begin{figure}[h]
\begin{center}
{\includegraphics[width=0.9\textwidth]{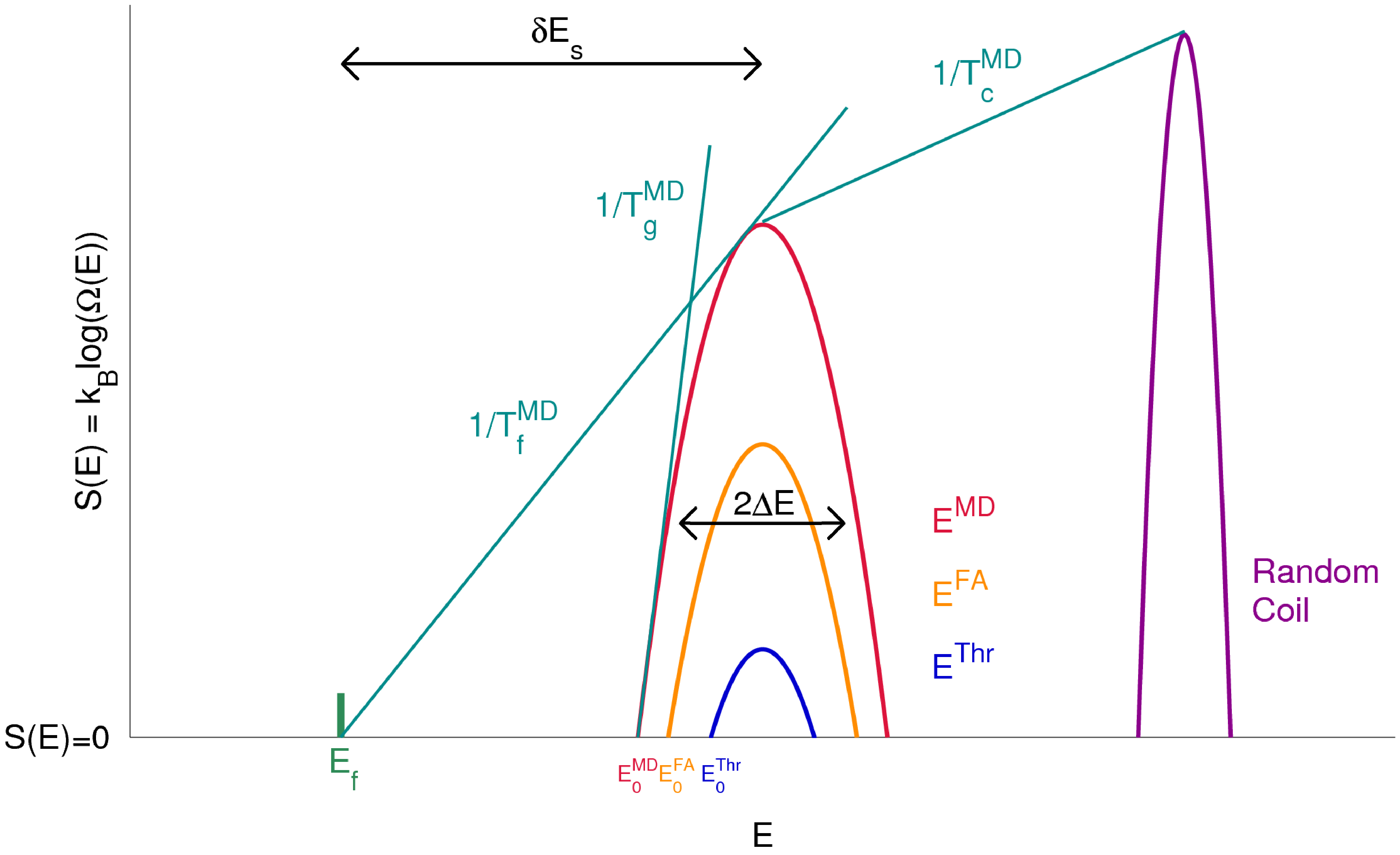}}
\caption{\label{fig:typesofdecoys} The theory behind this figure is described in Section~\ref{sec:theory}. A logarithmic plot of the number of structures with a given energy $E$. The expected ground state of a set of compact decoy structures corresponding to the molten globule can be inferred from landscape theory and is indicated by the intersection of a parabola with the abscissa. When there are more possible decoys the trap states become more competitive and are easier to confuse with the native state at $E_F$. The gap is reflected also in the characteristic temperatures $T_F$ and $T_G$ whose inverses are indicated as slopes on this diagram. A large $T_F/T_G$ corresponds with a large gap and easy recognition. Recognition becomes progressively more difficult as one moves from threading-based decoys (``Thr''), to fragment assembly (``FA'') and finally to fully flexible backbone molecular dynamics (``MD'').}
\end{center}
\end{figure}

In this article, we will mostly focus on structure prediction schemes of the latter sort that start by being based on an energy function and then use a wide-ranging search/sampling procedure. For a fully transferable method, the energy function and search must be able not only to predict the structures of the proteins on which it was trained (the training set) but also predict structures for proteins outside of the training set (the test set). If an algorithm produces more or less unambiguous, low energy, native-like structures for a wide variety of proteins upon minimizing an energy function, then this algorithm can be deemed a successful protein structure prediction scheme. The rest of this article will discuss how to design, optimize, refine and evaluate tertiary structure prediction methods using algorithms and ideas from the statistical energy landscape theory of protein folding. These considerations will be put into context by reviewing the historical progress of structure prediction using the AMH/AMC/AMW/AWSEM family of models. We will also discuss recent applications of these models to problems of finding binding sites for protein-protein recognition, multimer structures and characterizing misfolded protein structures that may be involved in protein folding diseases.

\clearpage

\section{Perspectives and assumptions}
\label{sec:perspectives}
How has nature solved the protein folding problem? The spontaneous folding of monomeric globular proteins is arguably the simplest kind of biological self organization. Folding generally involves only one molecule at a time, working, at least in most cases, without the aid of any other molecular actors except a suitable solvent. So no fancy biology needs to be invoked - chaperones, which after all consume valuable ATP, are actually used quite sparingly {\it in vivo}. The classical experiments by Anfinsen~\cite{anfinsen1972studies} gave credence to the idea that globular protein folding, and therefore protein structure prediction, can be achieved by minimizing an appropriately chosen free energy function. It is an increasingly well supported empirical fact that proteins with metastable states that are comparable in energy to the native state are the exception, not the rule. There may be a few specific proteins that are metastable like the serpins, but the metastability has evolved for a particular functional purpose~\cite{ferreiro2013frustration,baker1994kinetics}. In contrast many functional RNAs are metastable so as to cut off their action after a timely response to a time varying signal, otherwise excess protein would be produced by translation. Some intrinsically disordered proteins only become ordered upon binding, whereas some others remain partially disordered even while functioning. The problem of ``structure prediction'', {\it i.e.} ensemble characterization, for the latter type of intrinsically disordered proteins is an interesting problem but will not be considered in the present review~\cite{eliezer2009biophysical,latzer2006simulation,weinkam2005funneled,weinkam2009electrostatic}. In the case of coarse-grained models, the Hamiltonian represents a free energy function not strictly an energy function. It depends on the protein chain coordinates. What should be the properties of such a free energy function that allow robust predictable folding? Abundant evidence has amassed indicating that the energy landscapes of globular proteins are funneled towards their native state, a fact that has come to be known as the Principle of Minimal Frustration~\cite{bryngelson1987spin}. The Principle of Minimal Frustration is a statement about the relative importance of the interactions present in the native state, the so-called ``native interactions'', versus the happen-stance random non-native interactions that might form in alternative conformations. If native interactions are on average sufficiently strong compared to the non-native interactions then the energy landscape of the protein will be smoothly funneled toward native-like configurations and at low temperature Brownian motion will lead to the folded state. Only a small subset of all possible protein sequences have landscapes that satisfy the minimal frustration constraint. Since these sequences have been selected by evolution, funneled landscapes contrast sharply with the rugged landscapes of typical random heteropolymers.  For most random heteropolymers the global ground state is nearly degenerate with other very different structures and is separated from these structures by high barriers.  One way of achieving robust structure prediction, then, is to mimic the funneled nature that has evolved for natural protein energy landscapes. Energy landscape theory gives mathematical definiteness to quantify the concept of minimal frustration and thus provide algorithms to learn energy functions starting from a database of known foldable protein structures and sequences. We describe this guiding strategy in the rest of this article. Pursuing this strategy leads to energy functions that are similar enough to the one that has been used by nature so that these energy functions can be used not only for structure prediction but also for exploring motions outside of the folded basin. Although, for a single given sequence, energy functions funneled to the native structure are not unique, requiring a transferable energy function that is flexibly applicable to many sequences to be funneled simultaneously for many proteins in a training set does constrain the parameters in such an energy function considerably. Such an energy function is more like nature's energy function than those that can currently be constructed from short distance molecular physics alone.

Like many worthwhile problems, protein structure prediction has required sustained effort over the course of decades, and perfect structure prediction has yet to be achieved. Energy landscape theory still therefore provides a framework under which structure prediction methods can continue to improve as our understanding of protein physics evolves, as the number of experimentally determined structures continues to increase, and as the available computational power grows. 

It is useful to think about the problem of structure prediction, or indeed protein folding in general, in terms of other well studied physical phenomena and in analogy to other problems in statistical physics. If one looks just at the start and end points of the problem, the amino acid sequence and the full three dimensional structure of a protein, structure prediction appears to be a translation problem, but not one that is as simple as translating the one dimensional DNA sequence into a one dimensional amino acid sequence. Instead, the input and output information are of fundamentally different kinds, having different dimensionality. The one dimensional amino acid sequences of proteins are exceedingly diverse and can appear almost random if analyzed naively~\cite{weiss2000information}, but the three dimensional structure corresponding to a given sequence appears to be nearly unique (at the resolution of crystallography, at any rate - we do not consider here the description of conformational substates lying within the folded basin~\cite{frauenfelder1991energy,fraser2011accessing}). Folding is therefore very much a problem of discrimination. The molecule must be able to discriminate between structurally distinct states, some of which would be nearly degenerate in energy if the amino acid sequence were truly random. The nearly unique folded state must be stabilized {\it and} the many possible misfolded states must all be simultaneously destabilized in order to prevent trapping during any search procedure.

\begin{figure}
\begin{center}
\label{fig:funnelfigure}
\includegraphics[width=0.5\textwidth]{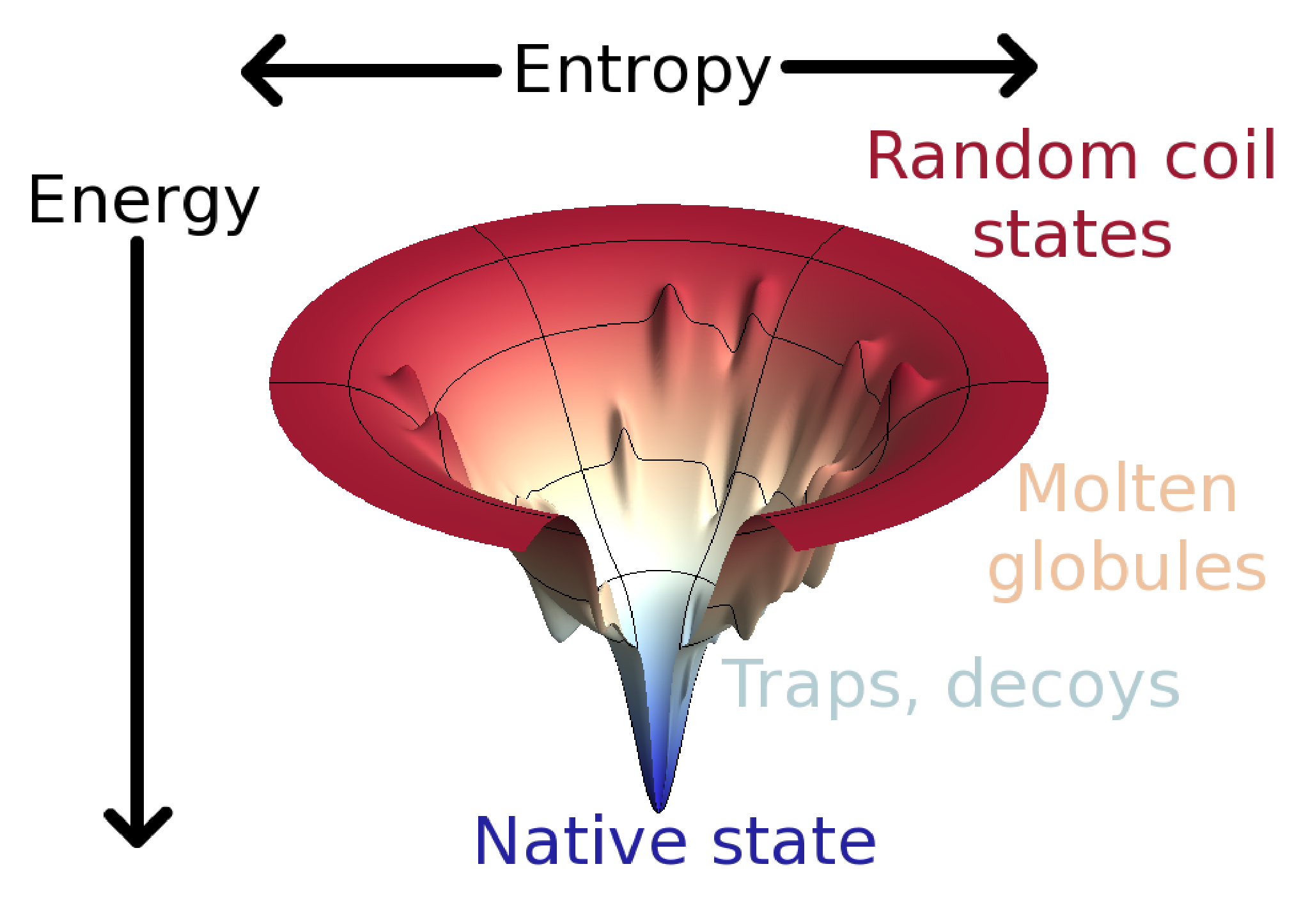}
\caption{Successful and efficient search is possible if the energy landscape is funneled. In this funnel diagram, the depth represents the solvent averaged free energy of specific structures while the width represents the entropy of possible states. This figure is essentially Figure~\ref{fig:typesofdecoys} turned on its side. Random coil states at the top of the funnel are like the gaseous state. Compact candidate structures are the so called molten globule or liquid phase and the worst traps or decoys are deep local minima that impede search and might be confused with the native state.}
\end{center}
\end{figure}

In many respects, then, folding resembles a nucleated, first-order like phase transition, a crystallization, but in a finite system. At the top of the funnel are many states with very few intrachain contacts, corresponding to a gas-like phase of a single protein. At the bottom of the funnel is the nearly unique native state and its related conformational substates. As the configurations go from being completely extended to being more native-like, they must collapse, and a liquid-like molten globule phase may exist. These phases are illustrated in the form of a two dimensional schematic funneled energy landscape in Figure~\ref{fig:funnelfigure}.
It is a remarkable empirical observation that most proteins seem to fold directly from the gas-like phase in a two state manner. It is likely that proteins have evolved to do so because search for the native state from within the molten globule state is relatively slow just as crystallization from a liquid can be impeded by a glass transition. It has also been suggested that the cooperative folding of natural proteins has evolved for an additional reason: non-cooperatively folding structure/sequence pairs may be selected against in order to avoid aggregating through partially folded intermediates~\cite{dumoulin2005reduced}.

\clearpage

\section{Using theory to guide structure prediction}
\label{sec:theory}
The high dimensional nature of proteins and their myriad-possible conformations invites a statistical description of their energy landscapes ~\cite{bryngelson1987spin,wolynes2005energy}. Frameworks for analyzing many body interacting Hamiltonian systems using statistical methods have been developed in several contexts, especially in the theory of spin glasses where no special symmetry characterizes the problem. Spin glasses, sets of randomly interacting spins, have quite simple interactions but nevertheless exhibit subtle phase transitions. Much is still debated about the details of these phase transitions and yet fairly simple approximations can be useful for understanding how they work.  These approximations also allow one to estimate characteristic quantities such as the ground state energy. In particular, the random energy approximation, the approximation that the energy of two different configurations of a system are always independent regardless of their structural overlap, allows for a simple estimation of phase transition temperatures such as the ordering (or folding, in proteins) transition temperature and glass transition temperature. Because proteins are partially hydrophobic polymers that fold in a polar solvent the collapse temperature $T_c$ is also of practical importance when designing protein structure prediction algorithms. Below the collapse temperature search slows owing both to excluded volume and trapping, and below the glass transition temperature, in the collapsed state, search becomes essentially arrested; it becomes necessary to unfold and just try again.

The folding transition temperature determines the temperature below which the dominant ensemble of structures switches from being an entropy dominated ensemble to an energetically favorable ensemble with a more restricted set of structures, the folded minimum along with its functional conformational substates. In order for a protein to be kinetically foldable, it is necessary for the folding temperature of the protein to be larger than its glass transition temperature, and the larger the ratio $T_f/T_g$ a protein has, the more easily it will fold over a wide range of temperatures. For a minimally frustrated protein, a protein with a large enough value of $T_f/T_g$, the folding time scales polynomially with the chain size, rather than exponentially as would be expected for a random heteropolymer with a low energy ground state that is a compromise between many frustrated interactions~\cite{wolynes1997folding}. This is the basic way in which the funneled nature of natural protein energy landscapes, having high $T_f/T_g$, ``solves'' the Levinthal Paradox: with a consistent bias towards the native state throughout the configuration space, it is simply not necessary to sample the entire configuration space to fold. The biases instead lead Brownian motions rather smoothly to the free energy minimum.  Likewise, it has also been shown that if the landscape is funneled, the accuracy necessary to predict the ground state does not scale with chain size ~\cite{pande1995accurate,pereira1996monte,pereira1997estimates} while it would if the landscape were random (as pointed out by Bryngelson~\cite{bryngelson1994potential}).  This observation from landscape theory gives hope to practitioners of coarse-grain modeling since it shows that near perfection in the force field is not necessary which would be the case if native structures energetically only won out by a few $k_BT$, as has sometimes been suggested.  Landscape theory provides the hope that practical folding might achieve reasonable results with simple models and finite structural databases from which to learn parameters.

Theory provides simple formulas for calculating thermodynamic quantities related to the folding, glass and collapse temperatures. The folding and glass temperatures can be understood using the ``simplest viable protein folding landscape'' picture described in Bryngelson {\it et al.}~\cite{bryngelson1995funnels,onuchic1997theory}.  One begins with two postulates: A) The energy landscapes of proteins are rugged because of the possibility of making inappropriate non-native contacts between residues but B) the Principle of Minimal Frustration allows the native contacts to be energetically differentiated from the non-native ones.  We make the approximation that energy of collapsed unfolded conformations can be crudely described by the random energy model (REM).  The REM approximation assumes the energies of any two unfolded but collapsed conformations are independent of each other, regardless of structural overlap.  Plotkin, Wang, and Wolynes have shown that correlations in the landscape can be accommodated for, but surprisingly, the characteristic temperatures from the REM approximation are not too bad~\cite{plotkin1996correlated,plotkin1997statistical}.  Given this assumption, the probability distribution of energies of the ensemble of unfolded states can be described by a Gaussian characterized by an average energy, $\overline{E}$ and variance $\Delta E^2$.
\begin{eqnarray}
P(E)dE = \frac{1}{\sqrt{2\pi\Delta E^2}}\exp\bigg({\frac{-(E-\overline{E})^2}{2\Delta E^2}}\bigg)dE
\end{eqnarray}
The unfolded state has an entropy 
\begin{eqnarray}
S_o = k_B\log\Omega_o
\end{eqnarray}
where $\Omega_o$ is the number of unfolded but collapsed configurations.  The density of conformational states then follows
\begin{eqnarray}
\Omega(E) = \Omega_oP(E)
\end{eqnarray}
The logarithm of this quantity is plotted in Figure~\ref{fig:typesofdecoys} and is essentially the entropy of the system.  The total entropy as a function of the energy is then 
\begin{eqnarray}
S(E) \approx S_o - \frac{-k_B(E-\overline{E})^2}{2\Delta E^2}
\end{eqnarray}
where we have discarded the term that varies logarithmically with system size.  The most probable energy, $E_{mp}$ at a given temperature follows by finding the maximum of the thermally weighted canonical probability.  We thus maximize

\begin{eqnarray}
P_{th}(E) = \frac{\Omega(E)e^{-E/k_BT}}{Z}
\end{eqnarray}
which yields
\begin{eqnarray}
E_{mp} = \overline{E}-\frac{\Delta E^2}{k_BT}
\end{eqnarray}
On Figure~\ref{fig:typesofdecoys}, then, the most probable energy at a temperature T is found by drawing a tangent to the entropy curve with slope $1/T$.
The density of conformational states with energy $E_{mp}$ and the corresponding entropy for collapsed misfolded states are then given by
\begin{eqnarray}
\Omega(E_{mp})&=&\exp\bigg({\frac{S_o}{k_B}-\frac{\Delta E^2}{2(k_BT)^2}}\bigg)\\
S(E_{mp})&=&S_o - \frac{\Delta E^2}{2k_BT^2}
\end{eqnarray}
At a low enough temperature, the misfolded ensemble will experience an entropy crisis and thus undergoes a phase transition where non-native trapping would be inevitable. Setting $S(E_{mp})=0$ yields the glass transition temperature
\begin{equation}
T_g = \sqrt{\frac{\Delta E^2}{2k_B S_o}}
\end{equation}
We can see that the likely energy of the misfolded ground state is found when the count of states is near $0$, {\it i.e.} at this entropy crisis point.  Thus $E_0 = \overline{E} - \sqrt{\frac{2\Delta E^2S_o}{k_B}}$.  We can see from Figure~\ref{fig:typesofdecoys} that this ground state is deeper the more diverse the set of possible competing structures is, as measured by $S_o$. Recognition of the native state, therefore, becomes progressively more difficult when moving from the problem of threading, to fragment assembly and finally to fully flexible backbone molecular dynamics models, as is shown in Figure~\ref{fig:typesofdecoys}. 

We may write the free energy as
\begin{eqnarray}
F(T)&=&E_{mp}-TS(E_{mp})\\
F(T)&=&\overline{E} - \frac{\Delta E^2}{2k_BT} - TS_o
\end{eqnarray}
Since folding often follows two-state behavior, we define the folding temperature, $T_f$, as the temperature at which the free energy of the unfolded ensemble equals that of the native state for which we neglect the entropy. We then find at $T_f$
\begin{eqnarray}
F_{folded}&=&F_{unfolded}(T)\\
E_{folded}&=&\overline{E}_{unfolded} - \frac{\Delta E^2}{2k_BT_f} - T_fS_o
\end{eqnarray}
The stability gap between folded and compact unfolded states is $\delta E_s = E_{unfolded} - E_{folded}$ and it follows that
\begin{eqnarray}
T_f&=&\frac{\delta E_s + \sqrt{\delta E_s^2 - 2S_o\Delta E^2/k_B}}{2S_o}
\end{eqnarray}
For large $\delta E_s$, the expression simplifies to
\begin{eqnarray}
T_f&\approx& \frac{\delta E_s}{S_{o}} \nonumber \\
\end{eqnarray}

The collapse temperature also involves losing entropy due to excluded volume while gaining generic stability from hydrophobic contacts and from thermally selecting especially favorable random contacts ~\cite{sasai1990molecular, sasai1992unified}. An approximate formula for the collapse temperature in terms of the number of residues, $N$, and the energy gap between the random coil states and collapsed states $\delta E_c$ is given in Equation~\ref{eq:collapse}.
\begin{eqnarray}
T_c&\approx& \frac{\delta E_c}{Nk_B}
\label{eq:collapse}
\end{eqnarray}

While the REM approximation does not take into account correlations between different states, regardless of their structural overlap, a more detailed treatment, the generalized random energy model (GREM), which includes pair correlations, was originally proposed by Derrida~\cite{derrida1985generalization} and was reintroduced in the context of polymers and proteins by Plotkin {\it et al.}~\cite{plotkin1996correlated,plotkin1997statistical}.  They found that the thermodynamic quantities obtained from the free energy analysis are close to their REM values.  Similar results are found when the replica methods of spin glass theory are employed~\cite{shakhnovich1989formation, sasai1992unified}.  More subtle considerations regarding freezing at different length scales and the proportion of stabilization that comes from long range versus short range terms can also be addressed within simple theoretical frameworks and will be discussed further in the section on the details of the optimization of coarse-grained force field models.

\clearpage

\section{Choosing the form of a coarse-grained Hamiltonian}
\label{sec:hamiltonian}

\subsection{All-atom models versus template-free coarse-grained models}
When starting to build a model to perform protein structure prediction, it might initially seem appealing to include as many details of the structure and of the interactions as possible. Once the form of interatomic interactions has been determined, the only task left to perform would seem to be a tuning of parameters which dictate the relative strength of the interactions. Classical mechanical, pairwise additive, explicit solvent varieties of these ``all-atom'' models exist~\cite{brooks2009charmm,case2012amber}, and have recently found some success in folding small proteins~\cite{lindorff2011fast}. One of the main disadvantages of the full blown atomistic approach is, however, that because of the very many degrees of freedom and the roughness of fully atomistic energy landscapes, especially before they were properly tuned, the search for the global free energy minimum is computationally expensive. This has made the development of all-atom  force fields an arduous task, which has however met with success~\cite{lindorff2011fast,best2012optimization}.

The use of coarse-grained models~\cite{levitt1975computer,friedrichs1990molecular,liwo1999protein,kenzaki2011cafemol,davtyan2012awsem} can simultaneously ameliorate both of the problems that lead to difficult computations. By describing fewer degrees of freedom, the forces involved in a coarse-grained model are much faster to compute and by keeping only the important degrees of freedom many of the local minima are also eliminated. For example, the solvent minima corresponding to amorphous ice and ice clathrates are avoided. These are involved in many barriers in real folding. Fortunately, so long as enough structural detail is kept, and an appropriate optimization procedure is performed, predictive, transferable coarse-grained models can be derived. Strategies for how to optimize and refine these types of models will be discussed in subsequent sections.

Most of the speedup achieved by coarse-grained models actually comes from integrating over the solvent degrees of freedom, not the missing protein atoms. This is reasonable because solvent motions are typically fast compared to protein backbone motions that involve crossing dihedral angle barriers, and those motions that are not relatively fast can be aliased onto the Hamiltonian that only explicitly depends on the coordinates of the backbone atoms. The Hamiltonian is therefore a solvent-averaged free energy function, not literally a potential energy function, and the amount of time computing forces is dramatically reduced by going to the coarse-grained description.

It is important to remember that when building coarse-grained models for the purpose of structure prediction alone you do not always have to follow all the rules that nature follows. For example, it may be useful when designing a Hamiltonian that performs structure prediction via simulated annealing from an extended conformation to intentionally lower the barriers for dihedral flips in order to allow the configurations to be explored more rapidly than actually would occur in nature. Likewise, drying effects that come from expelling solvent when preformed subunits approach each other can give large barriers that slow kinetics for some protein folding processes~\cite{cheung2002protein}. We can also take advantage of the robustness of the funnel in evolution: given that we know many sequences that fold into the same structure, imperfections in the optimized parameters trained on a finite set of proteins can be overcome by averaging the basic force field for a given sequence over the Hamiltonians of many sequences within a family of related proteins~\cite{keasar1997simultaneous,keasar1998coupling,hardin2003associative}.  By working through the development of coarse-grained models, and seeing what succeeds and what fails, it has also been possible to get a feel for what the most important aspects of protein physics are and what amount of information and detail is actually necessary to not only predict structure but also understand folding kinetic mechanisms where getting the barriers right is often more important.

As the number of experimentally resolved structures increases, the fraction of newly resolved structures with genuinely novel folds that have not been seen before has been decreasing. A higher and higher fraction of sequences are therefore becoming good candidates for homology modeling, and homology modeling will remain a most reliable way of performing structure prediction for most sequences for some time. Nevertheless, the kinds of physically motivated coarse-grained models discussed in this review are useful for more than just tertiary structure prediction. The ability of a coarse-grained model to perform structure prediction is an important benchmark of its adequacy, and can be taken as a sign that the model is realistic enough to be used for other purposes; several examples of such applications to mechanistic questions are given in Section~\ref{sec:results}. Many interesting molecular biological phenomena, especially at the cellular level, are still well beyond the reach of all-atom simulations in terms of their time and length scales; optimized coarse-grained models are thus likely to be useful for many years to come. The combination of coarse-grained and all-atom models has already proved useful in many structure prediction applications~\cite{rohl2003protein}.

\subsection{Tertiary interactions, non-additivity and cooperativity}
\begin{figure}[h]
\begin{center}
{\includegraphics[width=0.4\textwidth]{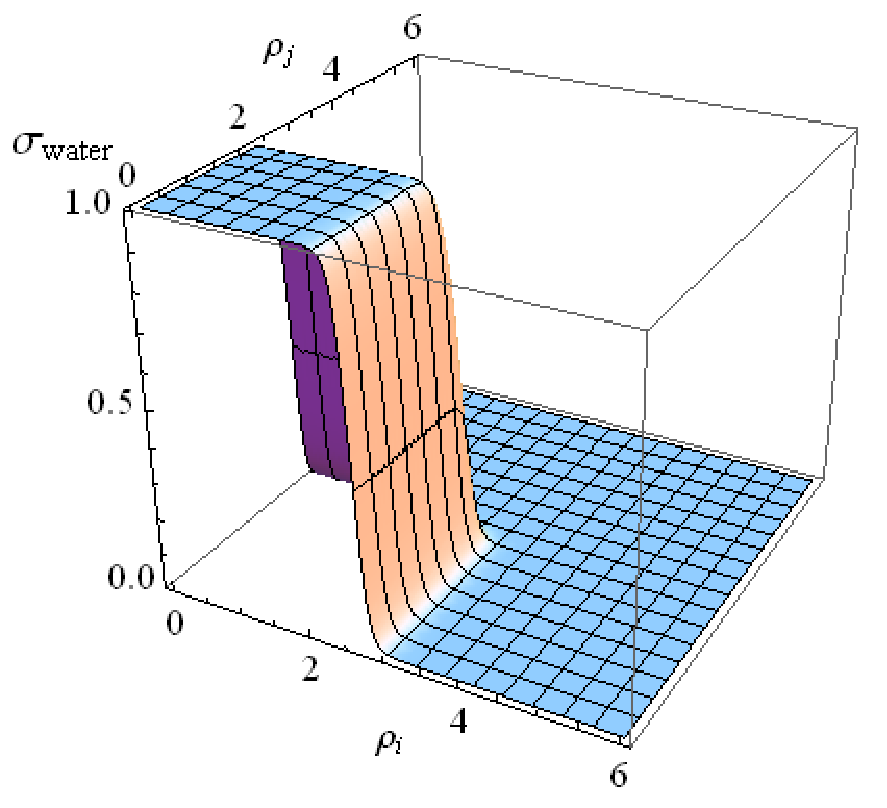}}
\caption{\label{fig:watermediatedinteraction} The water mediated interaction switches smoothly between two interaction weights depending on the degree of burial of the interacting residues. This switching function is shown as a function of the degree of burial of the two residues participating in the interaction. This figure was adapted from \cite{davtyan2012awsem}.}
\end{center}
\end{figure}

As more structural details of a protein model are integrated over, the appropriate form of the model energy function becomes increasingly less obvious. Building a coarse-grained model still retains elements of an art. Nonetheless, certain statistical mechanical principles are useful to keep in mind. For example, as a model becomes more and more coarse-grained, the pairwise additive approximation between degrees of freedom becomes less and less safe. It is therefore frequently useful explicitly to include contextual information about the local sequence and its environment to modulate otherwise pairwise additive interactions. An example from our own work is the water mediated interaction introduced in the AMW model. The motivation for implementing the water mediated interaction's particular functional form came from the desire to test whether the binding landscapes of hydrophilic and hydrophobic protein-protein interfaces were funneled or perhaps were rugged, leading to difficult binding search problems. Without the use of the water mediated interaction, even landscape optimized model energy functions that correctly predicted many monomer protein structures could only correctly predict the structure of hydrophobic interfaces~\cite{papoian2003role}. It was discovered that having water mediated interactions lead to better funneled folding landscapes~\cite{papoian2004water}, outside of the binding context in which it was originally motivated. The water mediated interaction is a sequence dependent pairwise contact interaction that switches smoothly between two different interaction weights depending upon the degree of burial of the interacting residues. It is therefore a non-additive potential. The switching function is illustrated in Figure~\ref{fig:watermediatedinteraction}.
If either of the two residues participating in the interaction are buried, the residues are assumed to be interacting indirectly through protein and that interaction is assigned a particular weight. If, on the other hand, both residues are exposed, the residues are assumed to be interacting indirectly through a water molecule, and are assigned a different interaction weight.

The water mediated interaction story illustrates the necessary interplay between observations, prediction quality analysis and implementation when developing coarse-grained models. The story also illustrates how the particular chosen functional form of a model determines the ultimate success of structure prediction. Part of the water mediated interaction that was introduced was a plausible one-residue burial propensity potential, another example of a non-pairwise additive potential, which sorts residues into their preferred burial environment (buried, partially buried or exposed). Residue-residue contact potentials and a single-residue burial propensity are two of the most commonly used styles of coarse-grained interactions. The motivation of this form is clear simply from looking at protein structures, where one finds {\it a la} Kauzmann, that certain residues prefer to be buried while others prefer to be exposed. At the next level of description, the pair level, it is clear that some residues prefer to be close to each other while others do not. Yet success requires the pair interactions to be modulated by burial, then leading to a very non-additive functional form. 

Protein folding of small globular domains is empirically a cooperative process, but subfolding events are also known to be cooperative. One example of a cooperative subfolding event is the formation of hydrogen bonds between two $\beta$ strands. This cooperativity can be explicitly introduced into coarse-grained models~\cite{kolinski1992discretized,hardin2003associative}. Explicitly introducing cooperativity into structure based models has been shown to be useful in achieving realistic barriers to folding~\cite{eastwood2001role} and in predicting and understanding hydrogen-exchange experiments~\cite{craig2011prediction}. Having realistically large barriers may actually hinder structure prediction schemes by impeding the search for low energy states. Nevertheless, achieving a realistic degree of cooperativity in predictive coarse-grained models remains an important challenge. Recent work indicates that the spatial range of the interaction potentials in coarse-grained models play a dominant role in determining the cooperativity of the model and that realistic cooperativity is obtained by making the interaction ranges consistent with desolvation physics~\cite{kaya2013spatial}.

\subsection{Backbone, steric and short range in sequence interactions}
The strongest constraints on protein structure are given by its very polymeric nature and by the specific chemical nature of its backbone which leads to stereo-typical possibilities of hydrogen bonding. Due to the Pauli exclusion principle, atoms essentially never overlap at biologically relevant energy and temperature scales, so it is important in coarse-grained models also to try to minimize the overlap that would occur if a higher resolution (all-atom) model were being used. Of course overlap at higher resolution cannot be entirely avoided when only a subset of the protein's atoms are considered, however. Thankfully, the local structures preferred by the protein backbone are surprisingly few. Most residues in natural protein structures can be unambiguously classified as being in either a helical or sheet conformation. In a coarse-grained model, these dominant local patterns can be satisfied by imposing a potential that acts on the dihedral angles of the backbone. Even when these constraints are taken into consideration, however, the specificity of the local conformation and how it relates to the local sequence is hard to capture in coarse-grained models due to the importance of local steric effects from overlap of sidechains with internal conformational freedom. These excluded volume effects are not explicitly included when the sidechain is treated with a unified representation. As a result, it is common to supplement the Hamiltonian for tertiary interactions with more information at the local-in-sequence level. Simply using fragments of possibly related experimentally determined structures is a popular method. Another method, used in AMW, is to perform bioinformatic alignments of global sequences to experimentally determined complete structures so as to enhance local compatibility in a mean field sense. This local information can then be used to impose gentle constraints on the relative distances of the backbone atoms nearby in sequence based on the analogous distances in the input candidate structures. Also one can use all-atom simulations of peptides to get structures that can determine the local biases~\cite{kwac2008protein}. Similar methods can be used to impose constraints on sidechain rotamer conformations.

The overall fold of a protein is determined by the trace of its backbone atoms. The trace of the backbone atoms is in turn determined by the secondary structural elements that form and the packing of these secondary structural elements. The representation of the backbone, therefore, is crucial to a good coarse-grained model of proteins. Ideally a backbone model restricts the backbone atoms to arranging themselves in realistic conformations while using a minimum number of degrees of freedom in order to keep computational efficiency. To this end, assumptions about bond lengths and planarity of the peptide bond are often employed, and dihedral angle potentials are used to mimic local steric effects. Steric effects in general can be an important factor in determining a coarse-grained model's ability to discriminate between what would be allowed and what would be disallowed configurations in a more detailed model. Usually coarse-grained models allow too many configurations that might lead to conflicts at an all-atom  level, and screening the results of coarse-grained models with more detailed models is a useful practice. Nevertheless, the steric constraints of all-atom  models can give an appearance of a much more rugged landscape than is correct simply because small flexible adjustments in all-atom conformation~\cite{mccammon1976hinge} can usually avoid the worst clashes.

\clearpage

\section{Inverse statistical mechanics: parameterization, optimization and refinement of prediction Hamiltonians}
\label{sec:optimization}
Rather than trying to compute the coarse-grained interactions starting from more basic molecular physics models or by using experimental constraints specific to a given protein as in structure based modeling~\cite{nymeyer1998folding,clementi2008coarse}, energy landscapes can be designed by the use of an inverse statistical mechanics approach to infer parameters. As we have seen, this approach does require some physical intuition in setting up the form of the Hamiltonian to be optimized and much has been learned through years of development. Certainly the interactions depend on the chemical identity of the interacting amino acid residues and thus having a residue-residue based contact interaction is an obvious but not unique choice of the form of the potential. While assigning parameters based on qualitative trends (e.g. the HP lattice code) may seem reasonable, the resulting models are usually not optimal in any quantitative sense. A better way of assigning parameters in an automated way is to use the quasi-chemical approximation that assumes pairs are assigned randomly by evolution.  This leads to the Miyazawa-Jernigan potential~\cite{Miyazawa1985Estimation}. In this approximation the strength of interaction between two residue types is proportional to the logarithm of the probability of finding those types adjacent to each other in the database of structures. The minimal frustration criterion on $T_f/T_g$ yields a related, but more sophisticated, and more optimal approach to assigning an ideal set of parameters.  The ingredients of the landscape optimization scheme we that we will describe requires the following: an energy function with parameters to be determined, a set of decoy structures which can be obtained in several ways, and a set of native structures, as a training set.

\subsection{Decoy structures}
The epigram at the beginning of this article is deliberately ambiguous.  It can be understood in multiple ways in the context of developing energy functions for protein structure prediction. First, through the continual development and refinement of energy functions for various structure prediction tasks, guided by the principles of energy landscape theory as well as by new experiments, misconceptions about protein physics and errors in the functional form of the potentials have turned into an increasingly coherent understanding of protein folding at the same time as more accurate and useful energy functions are developed. A more important lesson from the motto, however, is that the mistakes that have to be corrected when performing protein structure prediction correspond to the many possible misfolded configurations that must be simultaneously destabilized as the native state is preferentially stabilized. Generating mistakes is thus an important part of the landscape learning process.  Generating mistakes has not only been a problem for people making predictions but also for evolution which has solved the problem, through the trial and error of natural selection.  These misfolded configurations are sometimes called ``decoy structures''.  This term often brings to mind a small, fixed set while in fact the decoys span a cosmologically large space.  So the problem is how to use a small set to say something about the whole space: thus, statistical energy landscape theory.

Decoy structures can be generated in a number of ways, each of which has advantages and disadvantages. The simplest way of generating decoy structures is merely to shuffle the sequence of a protein while keeping its structure fixed. This would imitate the pairings in a highly mixed set of molten globule structures.  This is very computationally inexpensive but is unrealistic, particularly since there are, in reality, strong constraints on where particular types of amino acids can be placed. One such example is that of membrane proteins. In membrane proteins, the residues that reside in the hydrocarbon layer are almost completely hydrophobic, while the residues in the phosphate layer are enriched in polar and charged amino acids. Therefore, shuffling the sequence completely randomly creates unrealistic decoys that exaggerate the contribution of certain types of interactions to the stabilization of the native state (such as the polar-polar or charge-charge interactions, in this case). Another related way of generating decoy structures is to simply thread the native sequence, without shuffling it, but possibly allowing gaps, over a series of unrelated, known tertiary structures. Care must of course be taken to only thread over structures with at least as many amino acids as are in the native sequence, and this method is much better than simple shuffling. Both of these methods of finding a set of representative decoys still have the assumption that the set of misfolded structures is independent of the parameters of the Hamiltonian.  This is, of course, only an approximation, since energy landscapes based on pairs end up being correlated landscapes.  The ``right decoys'' must be found, leading to an iterative procedure.

\subsection{Self-consistent optimization}
The preferred method for generating decoy structures would be to explicitly generate those decoys that the Hamiltonian you are trying to optimize would actually have as kinetic traps in folding. While computationally expensive, requiring iteration to self-consistency, this repeated process of trial and error only needs to be done a few times for a given form of the Hamiltonian. Explicit decoy generation and self-consistent iteration is the best way to take into account the correlations that are present in the sequences and landscapes of natural proteins.  Thus, it is the best way of optimizing a set of parameters that will discriminate against the realistic decoy structures that would appear in a prediction attempt. Carrying out self-consistent optimization also requires an initial guess for the parameters in order to explicitly generate the decoys. Using shuffling or threading is a good way of generating an initial guess for the set of parameters in the Hamiltonian.

\subsection{Optimization functionals}
Given an (unparameterized) energy function, a set of decoy structures and a set of native structures, you might think that all that is needed is to be certain that the native fold is a bit more stable than the decoys.  This would be linear programming problem~\cite{maiorov1994learning}.  The problem is, however, you never have a complete set of decoys and to be transferable, the potential needs to yield a sizable energy gap.  One must be able to compute an average quantity that guarantees success with all possible decoys as competitors requiring using knowledge about the whole configuration space while knowing only a sample of the permissible decoy space.  Energy landscape theory fortunately tells us the right quantity that should be optimized in order to best discriminate between native and misfolded structures that will make the native state kinetically accessible. This quantity is the ratio $T_f/T_g$, or, equivalently, the ratio of the energy gap between the folded and misfolded states divided by the standard deviation of the misfolded energies $\delta E/\Delta E$.  If this parameter is large enough, the landscape is funneled and will provide a good thermodynamic discrimination between the misfolded and folded states at temperatures where the dynamics of search are still fast enough to access kinetically the native state (far from $T_g$). Simultaneously optimizing an appropriate average of the ratio of $T_f/T_g$ for a set of training proteins helps to ensure that the optimized parameters are as transferable as possible to proteins outside of the training set. This process requires some type of averaging over example proteins, which is to some extent arbitrary, but tricks such as weighting the contribution of each protein to the average in a way inversely proportional to its $T_f/T_g$ help to prevent the average from being dominated by only a few proteins with large $T_f/T_g$ and thus favors being able to fold the worst cases~\cite{vendruscolo2000comparison}. Even when a large training set is used, there is a large number of interaction parameters to be determined. Care, therefore, must also be taken to avoid noisy assignment of parameters. To prevent noisy interactions from dominating the contribution to the energy, a filtering scheme based on eigenvalue decomposition of the interaction matrices is used so that for examples of an interaction that is sampled poorly, their influence in the learning is minimized.  This is much like the way one tries to avoid learning superstitions by asking for robustness to coincidence.

\subsection{Constraints on optimization}
The generic polymeric nature of proteins which allows them to be collapsed or random coil, or microphase separated, for example, as well as specific structural peculiarities of proteins, make it useful to constrain other properties characterizing the landscape while optimizing the parameters to give funneled landscapes.  This leads to a constrained optimization problem. The additional constraints are necessary to control additional phase transitions such as the collapse transition.  These constraints enter into the optimization through Lagrange multipliers. Any physical consideration which can be expressed via summation over energy terms in the Hamiltonian, such as measures of local rigidity, collapse and contributions from various distance-range interactions, can be constrained in this manner. The degree of collapse is particularly relevant to control because the kinetics within a collapsed and misfolded ensemble is considerably slower from even just the steric constraints than it is within an expanded and also unfolded state, and of course, the larger the number of strong interactions that form the deeper the trap that results. When generating decoy structures, a generic collapse bias can therefore be applied to ensure sampling of the ``worst case'' - i.e., folding from the collapsed state. The preformation of secondary structure can lower the barrier to folding but also has the effect of slowing down rearrangements. There are simple physical arguments that suggest that the relative contributions of the local-in-sequence versus long range interactions are comparable~\cite{saven1996local}.  This idea can be used to constrain relative contributions of these terms in the potential parameters.  The variance of the local-in-sequence or long range interaction energy distributions also needs to be constrained so as to minimize the probability of there being a glass transition on short length scales before the global one occurs.  For example, a large variance in the local-in-sequence energy distributions may lead to dynamical freezing of those local-in-sequence interactions at $T > T_g$~\cite{plotkin1997statistical}.  Besides $T_f/T_g$ optimization, other related optimization functionals have been proposed and tested. Most of these include maximizing the free energy gap or native state occupancy in some way. In general, statistical landscape theory shows these measures of landscape funneledness are monotonically related to the $T_f/T_g$ criterion which we have generally employed.

\subsection{The mathematics of the optimization}
The mathematics of the optimization is simplest when the parameters that enter the energy function do so in a linear fashion, $E = \sum_i{\gamma_i\phi_i}$.  The $\gamma_i$'s are the strengths of the interaction terms whereas the $\phi_i$'s are monomials encoding the basic forms of the interaction  potential.  The stability gap can be written as $\delta E_s=\bm{A\gamma}$ whereas the energetic variance can be written as a general quadratic function $\Delta E^2=\bm{\gamma B \gamma}$.  $\bm{A}$ and $\bm{\gamma}$ are vectors of dimensionality equal to the number of interaction types while $\bm{B}$ is a matrix.  $\bm{A}$ and $\bm{B}$ are defined as\\
\begin{align*}
A_i&=\left<\phi_i\right>_{mg}-\phi_n\\
B_{i,j}&=\left<\phi_i\phi_j\right>_{mg}-\left<\phi_i\right>_{mg}\left<\phi_j\right>_{mg}
\end{align*}
where $\left<\phi_i\right>_{mg}$ and $\phi_n$ are, for a particular interaction type, the average $\phi_i$ of the decoy states and native state, respectively.  The optimization of $\delta E_s/\Delta E=\bm{A\gamma}/\sqrt{\bm{\gamma B \gamma}}$ with respect to the set of $\gamma_i$'s is equivalent to the maximization of $\bm{A\gamma}$ under the linear constraint that $\sqrt{\bm{\gamma B \gamma}}$ is constant.  That is, one optimizes with respect to the vector $\bm{\gamma}$ the functional, $R = \bm{A\gamma} - \lambda_1\sqrt{\bm{\gamma B \gamma}}$ where the Lagrange multiplier, $\lambda_1$, sets the energy scale. The solution of this geometric problem amounts to solving a system of linear equations $\bm{\gamma}=\bm{B^{-1}A}$ up to a scalar multiple. This is worked out in the Appendix.  We may also control the collapse temperature $T_c=\bm{A'\gamma}$ where $A'_i=\left<\phi_i\right>_{mg}/N$, the average $\phi_i$ of the decoy states divided by the number of residues in the protein, by imposing an additional constraint on our optimization functional.  This ensures the decoys to be generated by the Hamiltonian will have a similar degree of collapse.  Optimizing the new functional, $R = [\bm{A}-\lambda_2\bm{A'}]\gamma - \lambda_1\sqrt{\bm{\gamma B \gamma}}$ again yields a solution, $\bm{\gamma}=\bm{B^{-1}}[\bm{A}-\lambda_2\bm{A'}]$ up to a scalar multiple.  The Lagrange multiplier, $\lambda_2$ can be chosen to maintain the ratio of $T_f/T_c=1$.  The variance of the energy of the molten globules coming from different length scales can also be controlled by imposing an additional constraint on a new fluctuation matrix $\bm{B'}$ which is constructed using only the contributions from a particular length scale in the potential, such as the local-in-sequence interactions. The Lagrange multiplier constraining this new fluctuation matrix may be chosen to hold $\bm{\gamma B' \gamma}/\bm{\gamma B \gamma}$ constant.  The mean energy of the molten globules coming from different length scales can also be controlled by separating out the contributions from the different length scales in the term, $\bm{A'}$.  For the constraint, $\sum_{n=1}^k\lambda_k\bm{A'_k\gamma}$, where the index $n$ denotes the contributions from the $k$th length scale, the Lagrange multipliers $\lambda_k$ can be chosen to make the contributions equal.\\
For the simplest case, where $\bm{\gamma}=\bm{B^{-1}A}$ up to a scalar multiple, how is the optimization carried out in practice?  One begins with constructing a training set of experimentally determined native protein structures, taking into consideration factors such as sequence homology, sequence length, whether the proteins are globular or membrane proteins, and whether the proteins require cofactors.  Typically $\bm{A}$ and $\bm{B^{-1}}$ are computed using decoys generated via shuffling and averaged over the entire training set. The initial $\bm{\gamma}$ and resulting Hamiltonian is then used for explicit generation of decoy structures by molecular dynamics.  The decoys often include biasing to low $Q$ or constraints on the radius of gyration to ensure the configurations sampled have a similar degree of collapse to the native structure.  These explicitly generated decoys are then used to compute a new $\bm{\gamma}$, and the process is iterated until $\bm{\gamma}$ converges, thus the optimization is self consistent.  During each round of optimization, filtering of the $\bm{B}$ matrix may be required in order to minimize noise arising from poor statistics of certain types of interaction.  One recomputes $\bm{B^{-1}}$ by first computing an eigenvalue decomposition, $\bm{B}^{-1}=\bm{P\Lambda^{-1} P^{-1}}$ where the columns of the $\bm{P}$ matrix are eigenvectors and $\bm{\Lambda^{-1}}$ is the the inverse diagonal matrix of eigenvalues.  The contributions coming from unreliable eigenmodes are discarded by zeroing out the corresponding eigenvalues in $\bm{B^{-1}}$ below a cut-off.  One may also apply a damping condition when combining $\bm{\gamma}$'s between rounds of optimization in order to accelerate convergence, as is shown in Equation~\ref{eq:damping}.
\begin{equation}
\gamma'_{n+1} = (1-\alpha)\gamma_{n}+\alpha\gamma_{n+1}
\label{eq:damping}
\end{equation}

\subsection{Physical interpretation of optimized parameters}
\begin{figure}[h]
\begin{center}
{\includegraphics[width=0.9\textwidth]{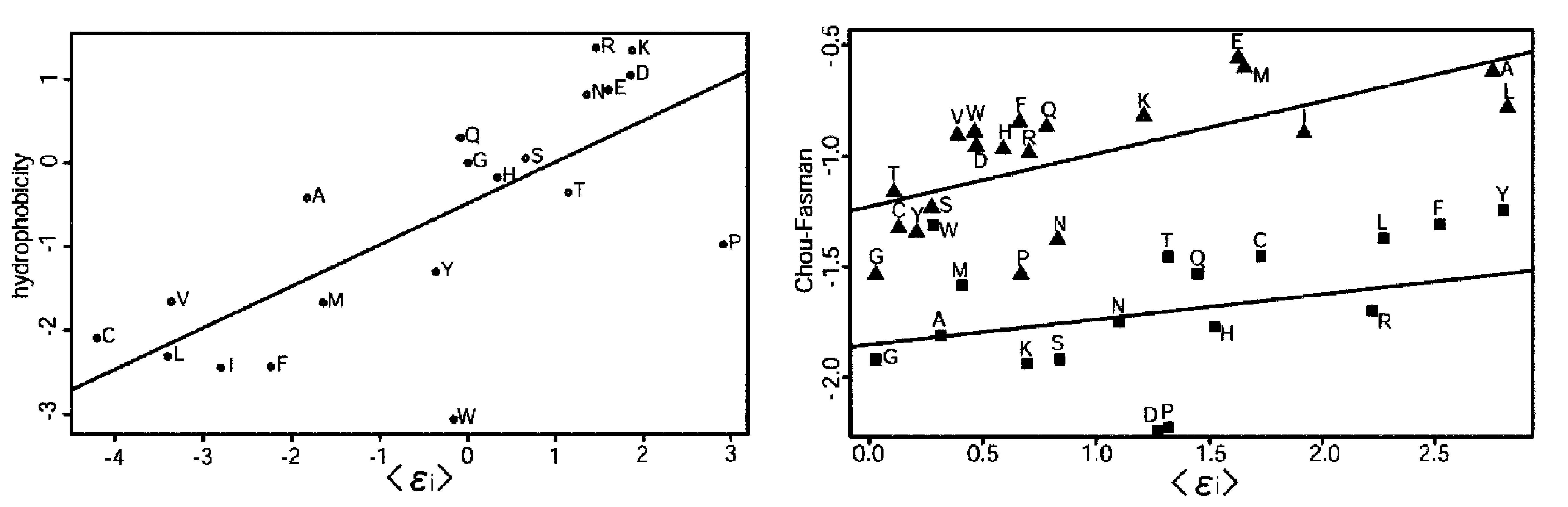}}
\caption{\label{fig:comparingphysicalandoptimizedparameters} Two plots showing the correlation between optimized parameters in a coarse-grained Hamiltonian and experimental quantities. The burial energy is correlated with a hydrophobicity scale and the secondary structure energies are correlated with experimentally determined secondary structure propensities. This figure was adapted from \cite{fujitsuka2004optimizing}.}
\end{center}
\end{figure}

Although no experimental information except the native structures of a limited training set has been used in parameterizing the Hamiltonian using a self-consistent $T_f/T_g$ optimization, the resulting parameters can sometimes be usefully compared to simple physical measurements, such as hydrophobicity and secondary structure propensity~\cite{fujitsuka2004optimizing}. An example of one such comparison is given in Figure~\ref{fig:comparingphysicalandoptimizedparameters}.  One can see in Figure~\ref{fig:comparingphysicalandoptimizedparameters} that the burial energy in the prediction Hamiltonian correlates with the hydrophobocity scale determined by transfer free energies from water while the secondary structure energy term correlates with the Chou-Fasman secondary structure propensity.

\begin{figure}[h]
\begin{center}
{\includegraphics[width=1.0\textwidth]{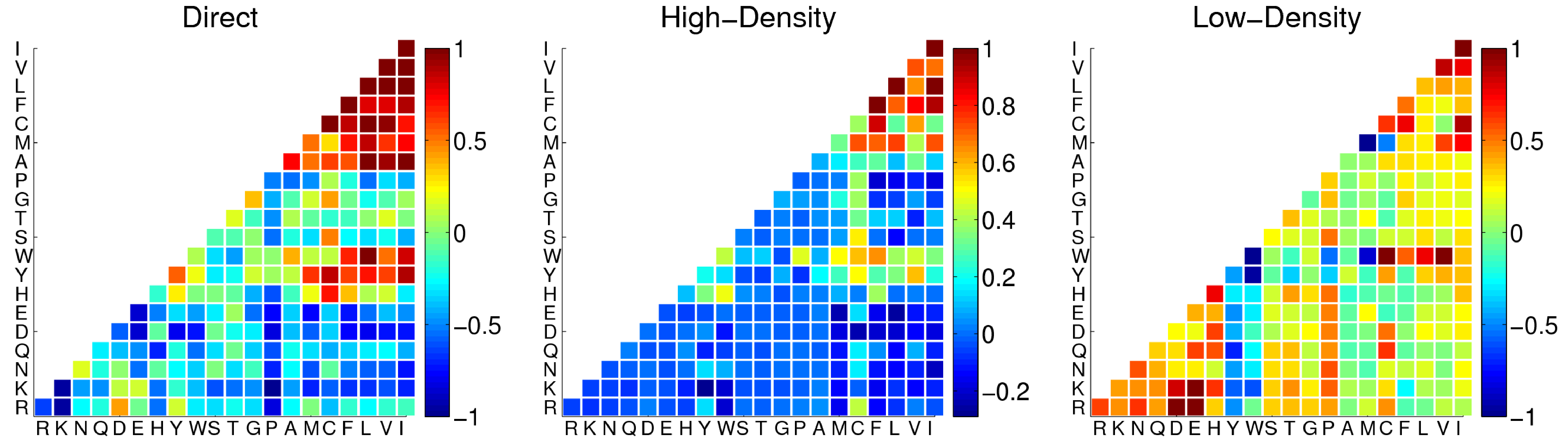}}
\caption{\label{fig:amwgammas} The three optimized interaction matrices used by the AMW/AWSEM models. The interaction weight for each pair of residue types (shown in one letter codes) is represented in color such that red interactions are favorable and blue interactions are unfavorable. Two residues interact using the interaction weights specified in the ``direct'' matrix when their $C_{\beta}$ atoms are between 4.5 and 6.5 $\AA\ $ of each other. When the $C_{\beta}$ atoms of two residues are between 6.5 and 9.5 $\AA\ $ of each other, the interaction can be either ``water mediated'' (Low-Density) or ``protein mediated'' (High-Density), depending on the degree of burial of the two residues.}
\end{center}
\end{figure}

The physico-chemical interpretation of the the AMW/AWSEM contact potential interaction parameters has been discussed previously in detail~\cite{papoian2003role, papoian2004water}.  The three interaction matrices utilized by the AWSEM contact potential are shown in Figure~\ref{fig:amwgammas}, where the more positive the interaction weight, the greater the energetic stabilization.  Polar interactions differ considerably among the different interaction types.  Differences in polar interactions between water mediated and direct contact strengths suggest a large desolvation penalty upon formation of direct contacts.  While polar residues are thus biased to water mediated contacts, the desolvation penalty, as expected, decreases the less polar the interaction.  Hydrophobic interactions are strongest for direct contacts, also as expected.

\begin{figure}[h]
\begin{center}
{\includegraphics[width=0.9\textwidth]{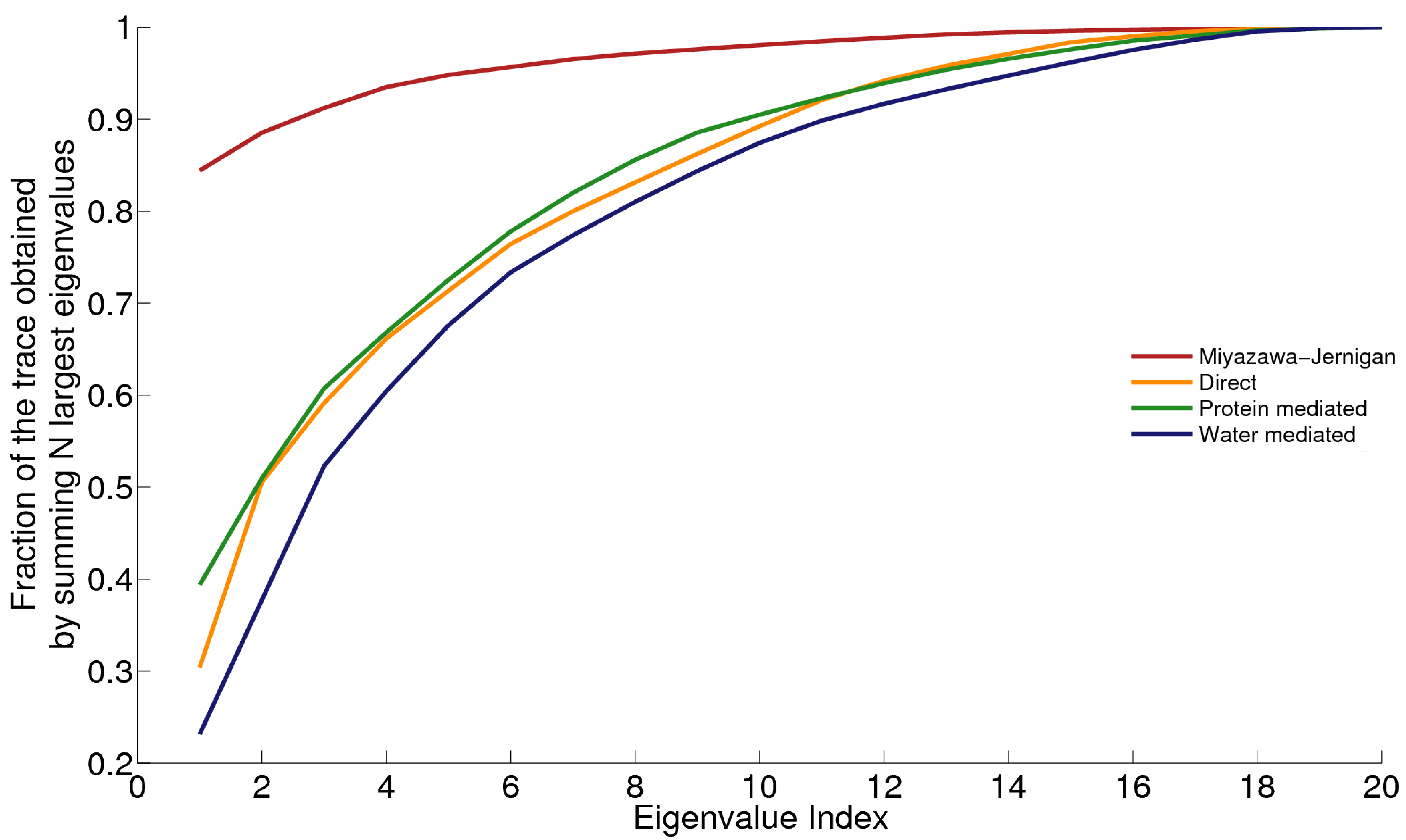}}
\caption{\label{fig:amwgammas_eigdecomp} The fraction of the trace obtained by summing the $N$ largest eigenvalues is plotted as a function of $N$ for four different interaction matrices including Miyazawa-Jernigan (red), and the three interaction matrices shown in Figure~\ref{fig:amwgammas} (orange, green and blue for direct, water mediated and protein mediated, respectively). The curve for the Miyazawa-Jernigan interaction matrix saturates much more quickly than any of the curves corresponding to the AMW/AWSEM matrices, indicating an overall lower information content.}
\end{center}
\end{figure}

We know the interaction matrices are not maximally complex from the general themes described about the physical chemistry of proteins already, but how complex is the protein folding code that comes from these optimized parameters?.  To assess the effective number of amino acid flavors the energy function encodes, one can inspect the eigenvalue decomposition of each individual interaction matrix, as suggested by Wingreen and coworkers~\cite{li1997nature}.  They found that the Miyazawa-Jernigan matrix could be accurately reconstructed using only the two largest eigenmodes corresponding to a hydrophobic-polar code, explaining why so much of a general nature about stability follows from hydrophobicity.  On the other hand, we find approximately 10 eigenmodes are necessary to reconstruct the interactions matrices employed by AWSEM as summarized in Figure~\ref{fig:amwgammas_eigdecomp}.  Clearly the forces involved in structure selection and specificity go far beyond the hydrophobic scale.

\subsection{Systematic refinement of parameters}
The optimization scheme described above is an efficient way of parameterizing linear parameters in a Hamiltonian. It is sometimes advantageous to also perform further refinements on parameters that enter in the Hamiltonian in a nonlinear way such as the ranges of the interactions once a reasonable set of initial parameters has been determined. One way of performing such optimizations is to simply scan through a set of parameters and perform simulations with each Hamiltonian. This, however, can be very computationally intensive. A faster way of reliably extrapolating the results from a simulation with one Hamiltonian to one using another energy function uses the idea of computing the properties of a perturbed Hamiltonian on a set of structures that were generated using an unperturbed Hamiltonian. The Free Energy Perturbation scheme of Zwanzig~\cite{zwanzig1954high}, and more general but related cumulant expansion methods~\cite{eastwood2002statistical}, provide systematic ways of extrapolating thermodynamic quantities. 

In some cases, the small perturbation idea is not valid - e.g., when adjusting the excluded volume radii of the particles in a simulation. For such cases, more sophisticated statistical mechanical methods can be used. The Mayer cluster expansion method is one such example~\cite{eastwood2003statistical}. Consistent with the optimization method described above, the Mayer cluster expansion method can be used to estimate how the folding and glass transition temperatures will change when the excluded volume parameters are changed in a residue specific way.

\subsection{Optimization and design}
It should not escape the notice of the readers that optimizing a Hamiltonian to fold a set of amino acid sequences into a given set of structures is very similar in spirit to the problem of designing a sequence to fold into a particular structure given a fixed Hamiltonian. The problems of design and optimization are dual to each other and much of the machinery described above can be and has been adapted to the protein design problem~\cite{samish2011theoretical}.

\clearpage

\section{Searching and sampling methods}
\label{sec:sampling}
We see then that structure prediction via protein folding simulation is a large scale optimization problem in two senses: first finding a free energy function and then sampling the low lying minima of that function. The free energy function once found still has to be minimized and it is high dimensional. When the energy landscape is funneled, finding the global minimum of the potential is no longer NP complete. Nevertheless, even on a well funneled landscape energy minimization and search can be computationally demanding for a big system.

The dynamics of the protein folding process is well described by low-dimensional diffusion equations, such as the one shown in Equation~\ref{eq:diffusion}. In Equation~\ref{eq:diffusion}, $P(Q,t)$ is the probability of being at a position $Q$ (some particular reaction coordinate such as $Q$ defined in Section~\ref{sec:evaluation}) at a time $t$, $D(Q,T)$ is a diffusion coefficient that could depend on $Q$ and temperature $T$, and $F(Q,T)$ is the free energy. The properties of the landscape enter in two gross ways into this equation, through $F(Q,T)$ and $D(Q,T)$. 
\begin{equation}
\label{eq:diffusion}
\frac{\partial P(Q,t)}{\partial t} = \frac{\partial}{\partial Q} \left\{ D(Q,T) \left[  \frac{\partial P(Q,t)}{\partial Q}+ P(Q,t)\frac{\partial \beta F(Q,T)}{\partial Q} \right] \right\}
\end{equation}
The approximate form for the diffusion coefficient is given in Equation~\ref{eq:diffusioncoefficient}. Search slows considerably when the sampling temperature becomes comparable to the roughness in the landscape, and it is therefore initially more favorable to sample at high $T$. 
\begin{equation}
\label{eq:diffusioncoefficient}
D(Q,T) = D_0\exp\left[-\Delta E(Q)^2/(k_BT)^2\right]
\end{equation}
However, thermal occupation of the native state is entropically disfavored at high $T$. For the purposes of structure prediction, then, good sampling techniques should either act on $F$ to lower or eliminate the free energy barrier to folding and/or act on $D$ to smooth the landscape by lowering or moving over local potential energy barriers and thereby minimize the folding time. 

\subsection{Simulated annealing}
Simulated annealing~\cite{press2007numerical}, the gradual cooling of a system from above to below its ordering transition temperature, has proven to be a general method for solving high dimensional optimization problems when the landscapes are funneled. Simulated annealing is typically carried out either via Monte Carlo moves or using molecular dynamics to explore configuration space. Molecular dynamics is straightforward to implement especially in parallel computer architectures but requires a Hamiltonian with continuous derivatives. Even when the the folding model is globally funneled, frustration in the native basin still can cause simulations to fall out of equilibrium at low temperature and thereby prevent the simulation from reaching the absolute lowest energy state~\cite{hardin2002folding}. If the landscape is sufficiently funneled, one nevertheless will still find a structure which is closely related to the true global minimum.

\subsection{Biased sampling techniques and equilibration}
Molecular dynamics combined with biased sampling techniques, such as umbrella sampling, can be used to obtain a global picture of the landscape and can be very helpful for assessing the structure prediction Hamiltonian itself and provides a way to assess the likely quality of the scheme's predictions, as will be discussed in Section~\ref{sec:evaluation}. Obtaining reliable equilibrium properties with finite sampling remains a problem, but techniques exist which can be used to test for equilibration. In one such method, samples from independent simulations are compared using the Kolmogorov-Smirnov test~\cite{eastwood2003statistical}.

\subsection{Flexible backbones versus fragment-assembly}
Coarse-grained molecular dynamics models with flexible backbones allow configurations that would be disallowed at higher resolution due to steric clashes. Fragment-based Monte Carlo methods benefit from having realistic backbone structures, and this often translates into high quality prediction results, particularly for smaller proteins. These methods have elements of {\it de novo} prediction at the global level mixed with homology modeling at the local level. The process of folding a protein is inherently coupled to collapse. Monte Carlo search, which requires large fragments to remain rigid during moves, suffers from an inability of simple moves to rearrange in the collapsed state resulting in an inability to avoid steric clashes~\cite{hegler2009restriction}. Then sampling even a well-funneled landscape in this way can be problematic for large structures if a simple move set is used. Direct fragment assembly also has issues with protein geometries that require small rearrangements such as $\beta$-strand alignment. The use of human insight in game playing has led to promising ways of avoiding these ultimately local dynamical problems~\cite{khatib2011algorithm}.

\subsection{Advanced sampling techniques}
Beyond molecular dynamics and Monte Carlo methods, many advanced sampling techniques exist. Methods such as basin hopping flatten the landscape making it easier to explore and find global energy minima, but this is done at the cost of breaking detailed balance~\cite{prentiss2008protein}. For this reason, such methods cannot be naively applied if the goal of the study is to obtain equilibrium properties, but these are a good choice if the goal is limited to finding the lowest energy states.  Advanced sampling methods for high-dimensional macromolecular systems were recently reviewed in~\cite{rohrdanz2013discovering}.

\clearpage

\section{How well do predictions work? Energy landscape analysis using order parameters}
\label{sec:evaluation}

The evaluation of different structure prediction methods is complicated by many factors including different test sets, training sets, model resolutions and potential uses of the model. Wherever possible subjectivity needs to be eliminated when evaluating structure prediction schemes~\cite{eastwood2001evaluating}. An objective criterion that can be used to compare and refine structure prediction methods is to quantify the degree of funneling, or how well the decrease in energy correlates with proximity to the native state. Other ways of assessing schemes, such as community-wide blind structure prediction competitions are also very useful~\cite{moult1995large}.

\subsection{Free energy and energy landscape analysis}
In order to obtain free energy profiles that can be used to objectively evaluate structure prediction Hamiltonians, it is necessary first to perform simulations biased by forces that constrain quantitative measures of the proximity of states to the known native structure to fixed values at multiple temperatures. These can then be combined and unbiased to obtain multidimensional free energy profiles using methods such as WHAM~\cite{kumar1992weighted} or MBAR~\cite{shirts2008statistically}. Biased simulations can be performed in many ways, and one popular method is umbrella sampling along a predefined reaction coordinate~\cite{torrie1977nonphysical,souaille2001extension}.

Free energy landscape analysis is useful because it gives an overview of which parts of the landscape will be sampled and which will not during prediction runs, or more specifically how long it will take to sample each part of the landscape, during an unbiased equilibrium simulation. Regions that are high in free energy will not be sampled very often at equilibrium; the system will spend most of its time in ensembles corresponding to low free energy regions. Given equilibrium free energy curves, it becomes possible to make quantitative estimates about how long it would take a model to find the native basin. It is also possible to make estimates of how many distinct structures exist at various points along the reaction coordinate~\cite{eastwood2001evaluating}.

In an ideally funneled landscape, such as those used in structure-based modeling, the native basin will be low in free energy below the folding temperature. The folding transition temperature itself is controlled by the interplay between entropy loss and energetic stabilization when going from the unfolded to the folded state. It is therefore useful to compare the free energy profiles obtained using structure-based perfect funnel models to those obtained using predictive transferable energy functions~\cite{eastwood2001evaluating}.

\begin{figure}[h]
\begin{center}
{\includegraphics[width=0.9\textwidth]{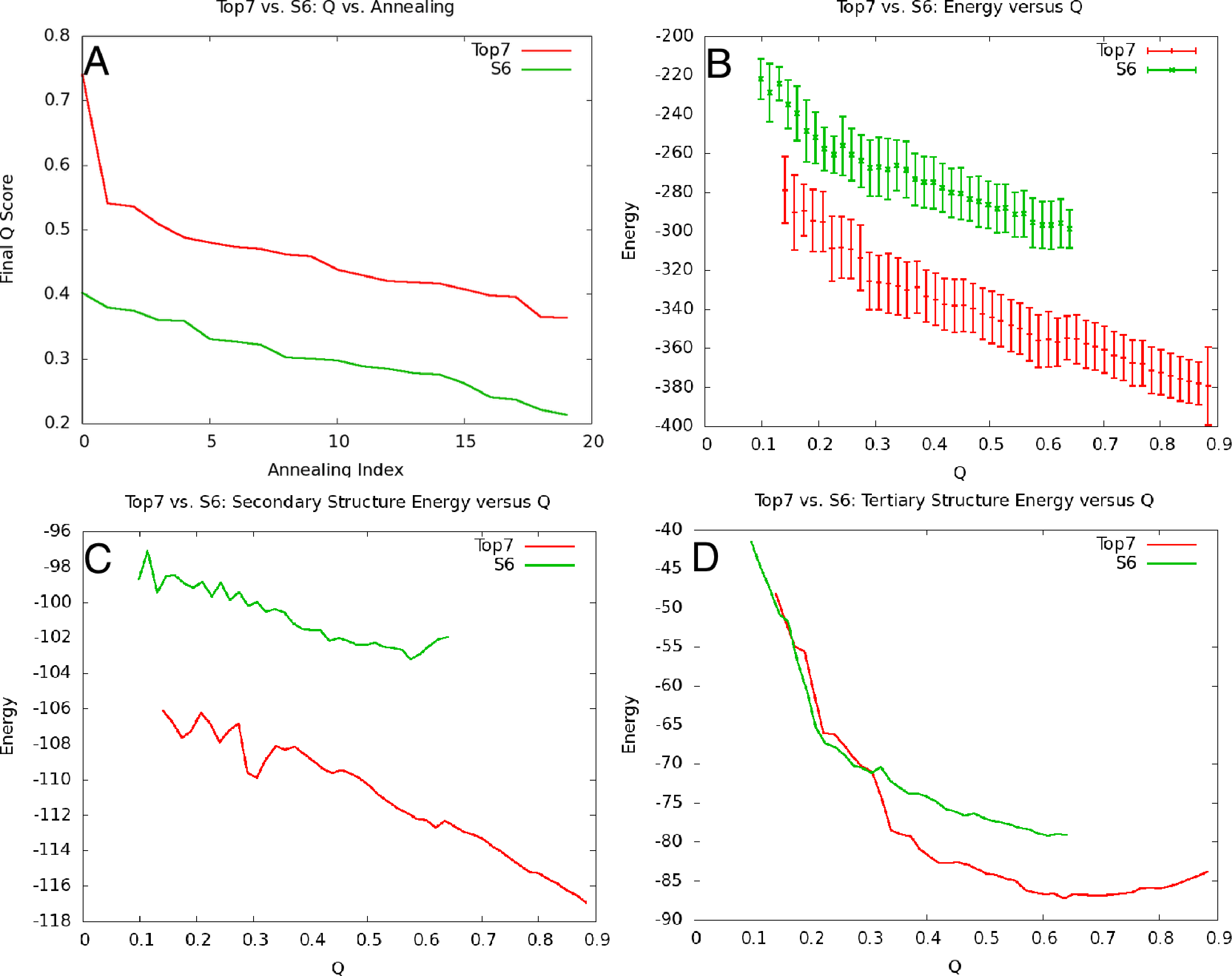}}
\caption{\label{fig:top7ands6energyanalysis} This figure is an example of how energy landscape analysis can be used to help understand structure prediction results. In (A), the results of the predictions of two proteins are shown by plotting the final $Q$ values obtained in 20 independent simulated annealing runs in order of decreasing $Q$. In parts (B), (C) and (D), the expectation values of the total energy, secondary structure energy, and tertiary structure energy are plotted as a function of $Q$, respectively. In this case, the desgined protein Top7 shows a greater degree of funneling to the native state than S6, and this is reflected in better structure prediction results. This figure was adapted from~\cite{truong2013funneling}.}
\end{center}
\end{figure}

Methods that allow for the calculation of free energy profiles can also be extended to the calculation of expectation values to characterize partially folded ensembles. One of the most interesting expectation values is that of energy versus degree of foldedness as measured by similarity of the contact map. If the expectation value of the energy versus similarity to the native state shows decreasing energy as the configurations become more native, and it follows that the gap between the native basin and unfolded basin is large in units of the variance of energies in the unfolded basin, the landscape is said to be funneled. Care must be taken in the interpretation of the results, however. Plots of $E(Q)$ are often funneled at high temperatures where the occupation of the native state is entropically disfavored. Therefore, the degree of funneling also must be checked at or below the folding transition temperature of the model, where the energetic stabilization in the native state does indeed overcome the entropic stabilization of the unfolded states. An example of energy landscape analysis in the context of a structure prediction application is shown in Figure~\ref{fig:top7ands6energyanalysis}. This example was taken from~\cite{truong2013funneling}, a structure prediction study comparing designed and natural proteins, which will be discussed further in Section~\ref{sec:results}.

\subsection{Choice of order parameters}
The choice of order parameters must also be done with care. Root-mean-square deviation (RMSD) is often used as a measure of distance to the native state, owing to its importance in x-ray crystallographic refinement. RMSD can be quite useful as a way comparing structures within the native basin. However, when looking at structures synoptically across the landscape, reaction coordinates like $Q$ are more useful because they are better correlated with the (largely contact-based) energy functions that are used. They also are more indicative of folding mechanism since contacts can form while the global topology is not yet set. The RMSD of structures with quite reasonable but partial contact maps can be very large. This may give the false impression that the folding landscape resembles a golf-course and is not funneled. The fact that $Q$ is unitless may make some people uneasy, but with a little practice it is easy to obtain an intuitive understanding of what different $Q$ values mean. Structures around $Q=0.25$ tend to be almost random looking, whereas $Q=0.4$ structures have reasonably well formed secondary structure and have significant topological similarity to the native state. At $Q=0.55$ the structure becomes easily recognizable and at $Q=0.7$ the structure is typically less than a few $ \AA\ $ RMSD from the native structure. This rough picture of the meaning of $Q$ is transferable across a range of protein sizes, which aids in the comparison of different systems.

When looking at global folding reactions, $Q$ has been found to be a useful reaction coordinate for kinetic analysis when the landscape is well funneled~\cite{cho2006p}. Likewise, when investigating specific phenomena it is often useful to create new reaction coordinates that are specific to the problem that you are investigating. For example, if you are interested in knowing how important short range in sequence interactions are in the folding, it makes sense to design additional reaction coordinates that single out the contribution to the folding from short range in sequence interactions alone. Also if a particular trap is important because of a specific frustrated part of the landscape, other coordinates may prove useful~\cite{sutto2007consequences}.

A generic formula for $Q$ is given in Equation~\ref{eq:q}.
\begin{equation}
\label{eq:q}
Q=\frac{1}{N^{\text{pairs}}}\sum_{ij\in \text{pairs}}\exp{\left[-\left(\frac{(r_{ij}-r_{ij}^N)^2}{2\sigma_{ij}^2}\right)\right]}
\end{equation}
In Equation~\ref{eq:q}, $N^{\text{pairs}}$ is a normalization factor equal to the number of terms in the sum. $r_{ij}$ is the instantaneous distance between atoms or groups $i$ and $j$, and $r_{ij}^N$ is this same distance but in the reference structure. $\sigma_{ij}$ is a (potentially sequence separation dependent) width of the Gaussian which sets the degree of tolerance to deviation from the reference structure and is typically on the order of an $\AA\ $. The list of pairs $\{ij\in \text{pairs}\}$ can be chosen to be either all possible pair distances, only those pairs in contact in the native state, or in other ways - such as those in specific foldon units~\cite{schafer2012discrete} - depending upon the application.

\clearpage

\section{The AMH/AMC/AMW/AWSEM family of models}
\label{sec:models}

\subsection{AMH}
\begin{figure}[h]
\begin{center}
{\includegraphics[width=0.5\textwidth]{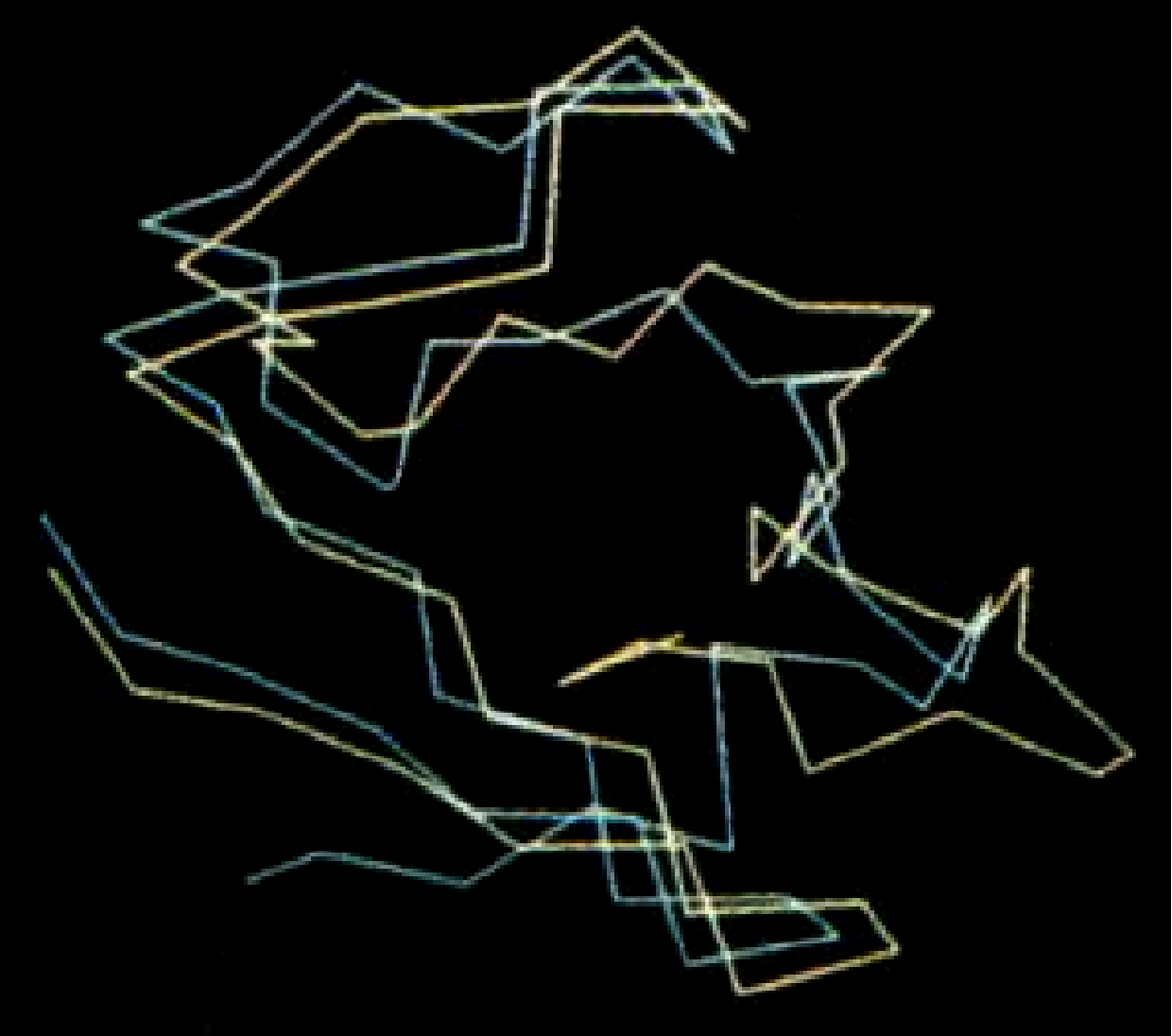}}
\caption{\label{fig:1989structure} A structure prediction result from 1989. This figure compares the predicted structure of {\it Desulfovibrio vulgaris} rubredoxin using an associative memory Hamiltonian containing 80 possible structures, only one of which was the homolog from {\it Clostridium pasteurianum} which differs from the {\it vulgaris} sequence in 50\% of its positions, demonstrating the model can generalize at least to the extent of local mutational substitution. The search algorithm employed a Monte Carlo assignment of local dipeptides from a database of known structures~\cite{friedrichs1989toward}.}
\end{center}
\end{figure}

\begin{figure}[h]
\begin{center}
{\includegraphics[width=0.5\textwidth]{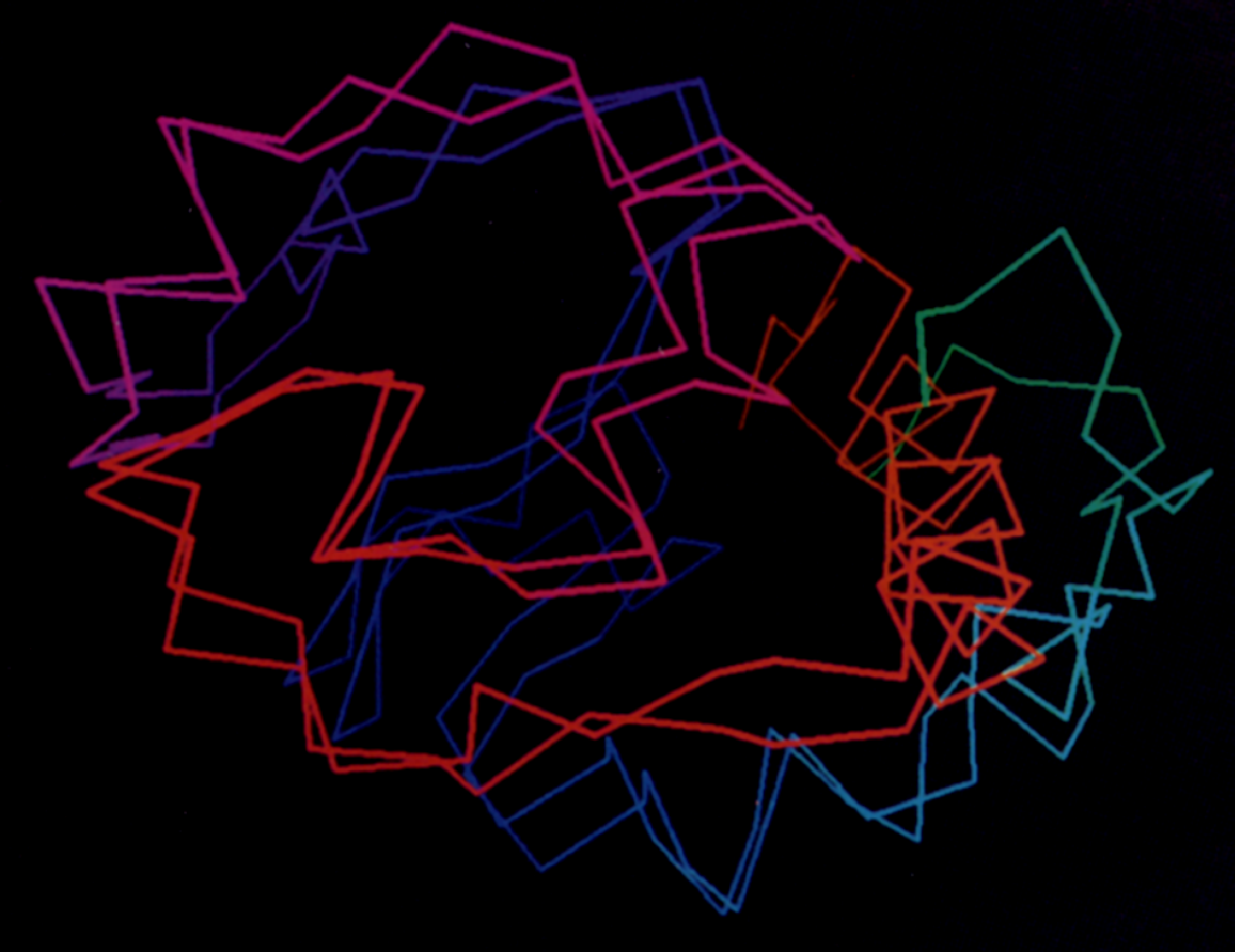}}
\caption{\label{fig:1991structure} A structure prediction result for cytochrome C from 1991. The unoptimized associative memory Hamiltonian used a set of memories that included other cytochromes but these had significant insertions and deletions in their sequence. A molecular dynamics based annealing method was used~\cite{friedrichs1991generalized}.}
\end{center}
\end{figure}

\begin{figure}[h]
\begin{center}
{\includegraphics[width=0.5\textwidth]{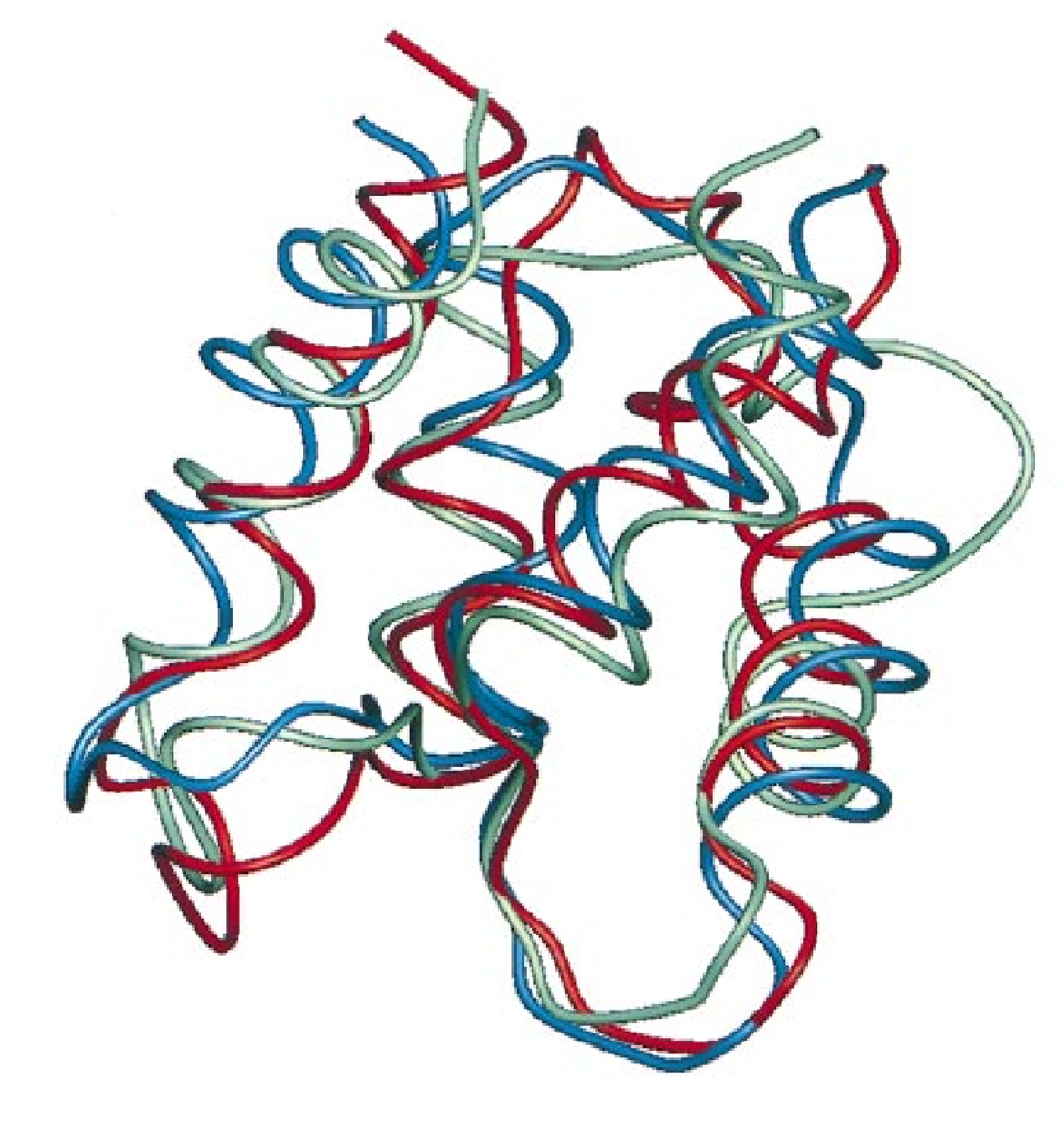}}
\caption{\label{fig:1998structure} A structure prediction result from 1998 for a calcium binding protein using an optimized associative memory Hamiltonian. While the memory set contained homologs, they were distance in sequence identity and the final structure showed the algorithm to be ``creative'' in that it was closer to the native structure than any input homolog. The correct structure is shown in green, the prediction in red, and the best input homolog in blue~\cite{koretke1998self}.}
\end{center}
\end{figure}

The AMH/AMC/AMW/AWSEM family of models is a series of coarse-grained protein folding models that have been continually developed, mostly in the Wolynes group, more recently with Clementi and in the Papoian group, over the last 24 years. The original version of the model, the Associative Memory Hamiltonian (AMH), was motivated by the neural network models of Hopfield and Little~\cite{hopfield1982neural,little1975statistical}. In the AMH model~\cite{friedrichs1989toward}, different residues have ``charges'', and interactions between residues depend on the value of these charges. In early papers, these charges were empirically defined using concepts such as hydrophobic vs. hydrophilic tendencies of the residues. Much as the way spin models were setup to ``recall'' a particular configuration given a different configuration that was nearby in configuration space, these early models were able to ``recall'' the native structure from a database of input structures given the sequence or a closely similar one as input. An example prediction using this model is shown in Figure~\ref{fig:1989structure}. It was found using analytical theory and confirmed in simulations that beyond a certain number of candidate structures (the ``capacity''~\cite{friedrichs1990molecular}), the simplest associative memory model becomes unable to faithfully recall the native structure given an input sequence. Such multiple memory models are useful for describing allostery~\cite{okazaki2006multiple,okazaki2008dynamic,itoh2010entropic,li2011frustration}, however. It was later found that, in some cases, these types of models could ``generalize'', {\it i.e.}, could produce predicted structures that were closer to the native structure of the input sequence than any of the homologs in the database of input structures~\cite{friedrichs1991generalized}. The ability to generalize was found to depend critically upon the correct choice of charges. In this way, generalization is closely related to the problem of finding the symmetries between different amino acids that are allowed by evolution. Grouping sequences by assigning similar values of the charge to ``biologically symmetric'' residue types effectively increases the number of sequences for which the structure can be reliably predicted.  Successful predictions are shown in Figures~\ref{fig:1991structure} and \ref{fig:1998structure}. However, simple intuition about what the correct choice of charges proved to be insufficient, and so a systematic way of optimizing parameters in these types of coarse-grained models of proteins was implemented.

\clearpage

\subsection{AMC}
\begin{figure}[h]
\begin{center}
{\includegraphics[width=0.5\textwidth]{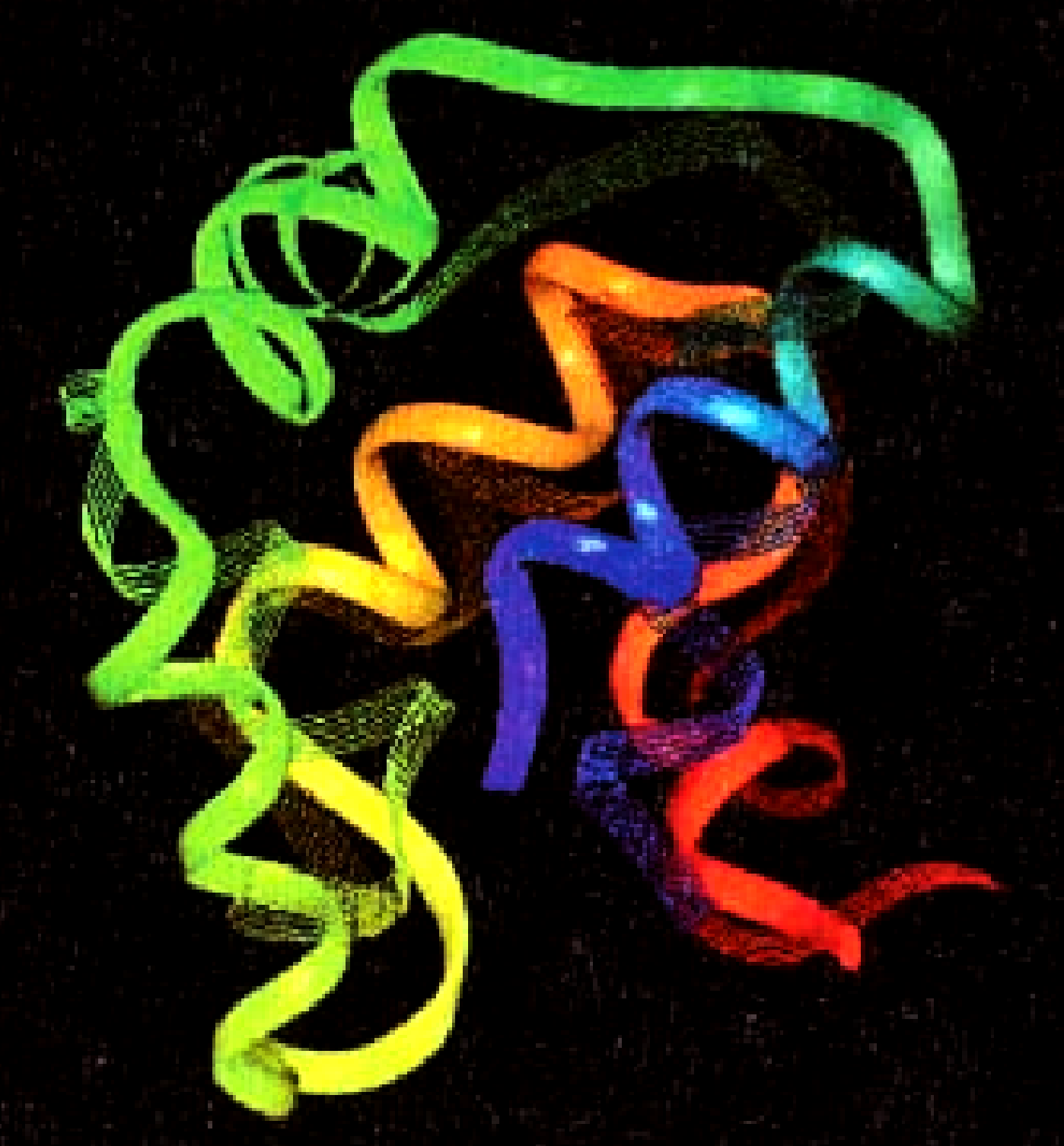}}
\caption{\label{fig:2000structure} A structure prediction result from 2000. A comparison of predicted and actual structures of 434 repressor. This result was obtained with an associative memory Hamiltonian having an optimized contact Hamiltonian. No homologs were used in the memory set~\cite{hardin2000associative}.}
\end{center}
\end{figure}

\begin{figure}[h]
\begin{center}
{\includegraphics[width=0.5\textwidth]{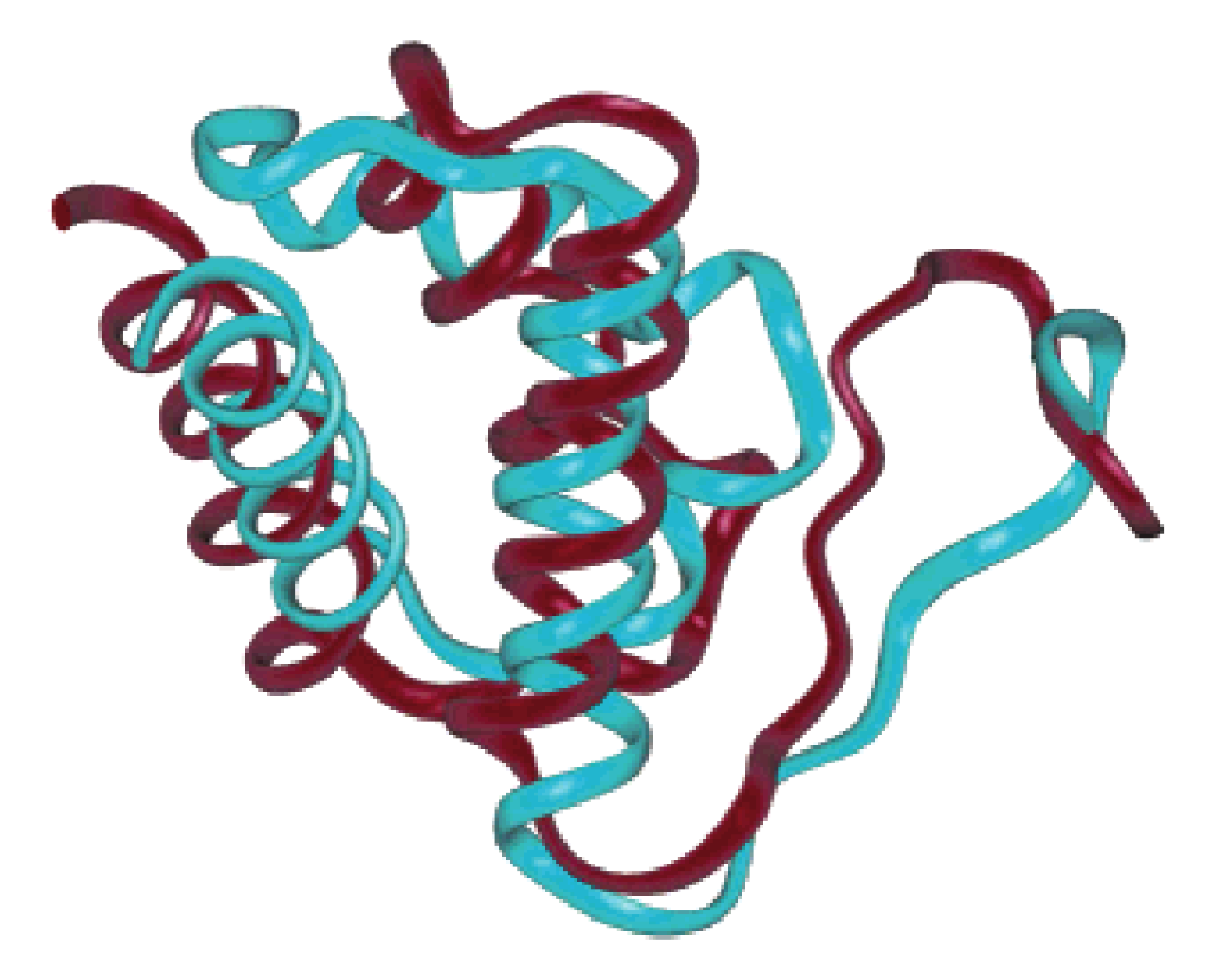}}
\caption{\label{fig:2006structure} This was the result of a blind prediction made by the Wolynes group of CASP5 target T0170 with current PDB code 1UZC. This was one of the best models for this fold submitted in that round of CASP. The methodology was based on the AMC code~\cite{prentiss2006protein}.}
\end{center}
\end{figure}

As discussed in Sections~\ref{sec:theory} and \ref{sec:optimization}, energy landscape theory can be used to optimize parameters~\cite{goldstein1992optimal}. This was first done in the context of the AMH model to assign weights to different interactions coming from memories. In particular, self-consistent optimization of the Associative Memory Hamiltonian using $T_f/T_g$ as the optimization function was shown to yield quantitatively correct structures. Later, the functional form of the potential was modified so as to use the Associative Memory interactions only for the short-range in sequence interactions while using a contact interaction for long-range in sequence interactions (Associative Memory with Contact, AMC~\cite{goldstein1992protein}). This allows generalization to include arbitrary length insertions or deletions. The contact energy based model was reoptimized using a similar scheme (also based on an optimized local energy function for the initial threading to find memories~\cite{koretke1996self,koretke1998self}) and was successfully applied to the problem of $\alpha$-helical protein structure prediction without the use of any homologs in the Associative Memory database~\cite{hardin2000associative} documenting then ``{\it de novo}'' structure prediction capability. Two AMC structure predictions are shown in Figures~\ref{fig:2000structure} and \ref{fig:2006structure}. The original version of the AMC model had three contact wells extending out as far as $15 \AA\ $ and used a global alignment scheme to obtain lists of ``memories'' for the Associative Memory interactions. Further refinements were added later including the addition of an explicit $\beta$-hydrogen bonding potential with cooperativity between nearby hydrogen bonds and a refined Ramachandran potential that could be further biased based on secondary structure predictions to obtain more realistic distributions of the backbone dihedral angles~\cite{hardin2002folding}.

\clearpage

\subsection{AMW}
\begin{figure}[h]
\begin{center}
{\includegraphics[width=0.5\textwidth]{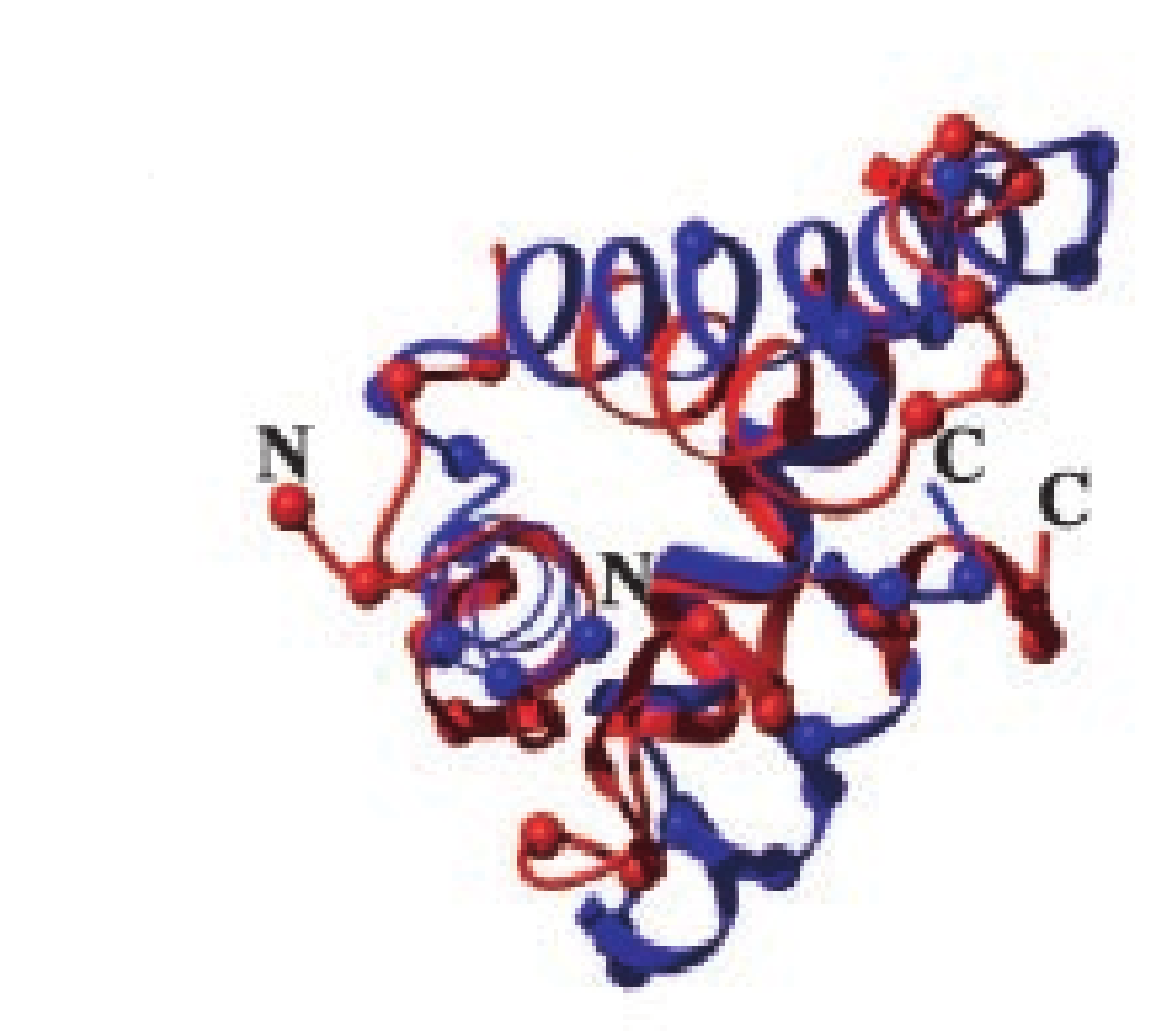}}
\caption{\label{fig:2004structure} A prediction of the CASP3 target HdeA. This prediction employed the optimized water mediated interactions~\cite{papoian2004water}.}
\end{center}
\end{figure}

The next major innovation came in new physics the form of the introduction of explicit water mediated interactions (Associative Memory with Water Mediated Interactions, AMW). The necessity of using the water mediated interaction emerged from a study of dimeric interfaces by Papoian and Wolynes~\cite{papoian2003role}. They showed that a simple pairwise additive contact potential proved insufficient for recognizing dimeric interfaces when those interfaces contained a significant amount of water. Knowing the problem - in a sense learning from errors - the functional form of the potential could then be modified to allow for the possibility of two residues interacting with different interaction weights depending on the local density of residues around each of them, {\it i.e.}, whether or not the interacting residues are buried or exposed. If both residues are exposed, and they are separated by a distance that is sufficient to allow a water molecule to fit between them, then they are said to be participating in a water mediated interaction. Once the model was reoptimized in this form, it was able to recognize both dry and wet interfaces. There were some surprises in the resulting potential. For example, it turns out that oppositely charged residues sometimes have an effectively favorable interaction because they are interacting with the same water molecule or perhaps an ion from the solvent. The AMW model was later used to predict the structures of monomeric proteins and the same water mediated interactions which allowed wet interfaces in dimers to be recognized were shown to also be important in the way larger proteins fold~\cite{papoian2004water}. Figure~\ref{fig:2004structure} shows a prediction on an early CASP competition target. 

\begin{figure}[h]
\begin{center}
{\includegraphics[width=0.5\textwidth]{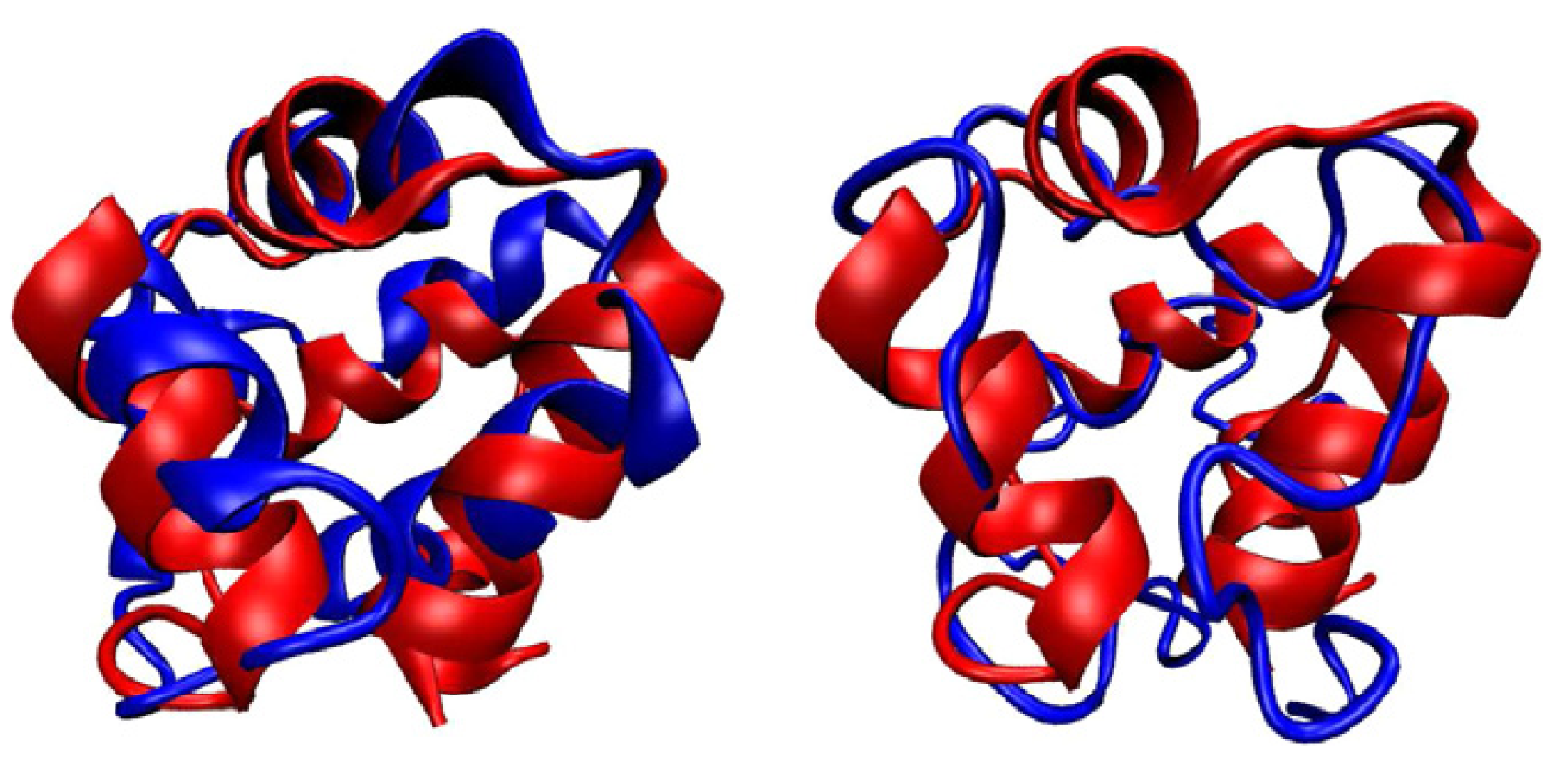}}
\caption{\label{fig:2010structure} This figure shows the quality of prediction for 1r69 that can be obtained from an optimized potential containing water mediated interactions and optimized hydrogen bonding energies alone. No local bioinformatic input was used. The left-hand structure use a generically predicted secondary structure based on propensities as input, the right hand prediction used no secondary structure bias at all. Predicted structure traces are in blue, the crystal structure in red. Clearly local signals are helpful~\cite{oklejas2010protein}.}
\end{center}
\end{figure}

To see whether a completely physically motivated model lacking any local structural biases as input could be used to determine local structures, a model having only water mediated interactions along with an $\alpha$-helical hydrogen bonding terms was also used to predict the structure of $\alpha$-helical proteins~\cite{oklejas2010protein}. While some results from this pure physics based model are fine, in general the prediction results obtained were significantly worse than predictions that also used bioinformatic alignments to determine associative memory forces to guide the formation of local structures. The difference between using secondary structure prediction information and not using any secondary structure input can be seen clearly in Figure~\ref{fig:2010structure}.

\clearpage

\subsection{AWSEM}
There are many ways to choose the short range guiding forces: all-atom physical simulations of fragments~\cite{kwac2008protein}, global sequence threading onto templates~\cite{koretke1996self}, local sequence similarity of peptide fragments~\cite{hegler2009restriction,davtyan2012awsem}. The associative memory, water mediated, structure and energy model, or AWSEM, is a concrete open-source instantiation of the AMW model. It uses rather simple direct bioinformatic searches to identify local fragments to use as memories and uses these memories to dictate the short range in sequence interactions. Otherwise, the interactions are made up of a combination of contact terms including the water mediated interactions and $\alpha$ and $\beta$ hydrogen bonding potentials. The primary advantage of AWSEM over previous instantiations of the model is that AWSEM is integrated into the LAMMPS molecular dynamics package~\cite{plimpton1995fast} and is therefore fully open source. A summary of the Hamiltonian is given in Equation~\ref{eq:awsem}.
\begin{eqnarray}
\label{eq:awsem}
V_{AWSEM} &=& V_{backbone} + V_{contact} + V_{burial} + V_{HB} + V_{FM} \\
V_{backbone} &=& V_{con} + V_{chain} + V_{\chi} + V_{rama} + V_{excl}
\end{eqnarray}
In Equation~\ref{eq:awsem}, $V_{backbone}$ consists of several terms which are responsible for maintaining the connectivity of the polymeric peptide backbone, ensuring correct chirality of the amino acids, restricting the conformations to reasonable dihedral angles and preventing overlap of the chain with itself. $V_{contact}$ consists of a direct, pairwise-additive contact term as well as the nonpairwise-additive water mediated interaction discussed previously. $V_{burial}$ sorts amino acid types into their preferred burial environments. $V_{HB}$ is made up of both $\alpha$-helical and $\beta$-strand hydrogen bonding terms. $V_{FM}$ uses information from bioinformatic alignments to bias local-in-sequence configurations. A complete description of the model is given in the Supplementary Information of a recent paper by Davtyan {\it et al.}~\cite{davtyan2012awsem}.

\clearpage

\section{Recent results}
\label{sec:results}

\subsection{Prediction of monomeric protein structures using simulated annealing of AWSEM with and without the use of homology}
\begin{figure}[h]
\begin{center}
{\includegraphics[width=0.9\textwidth]{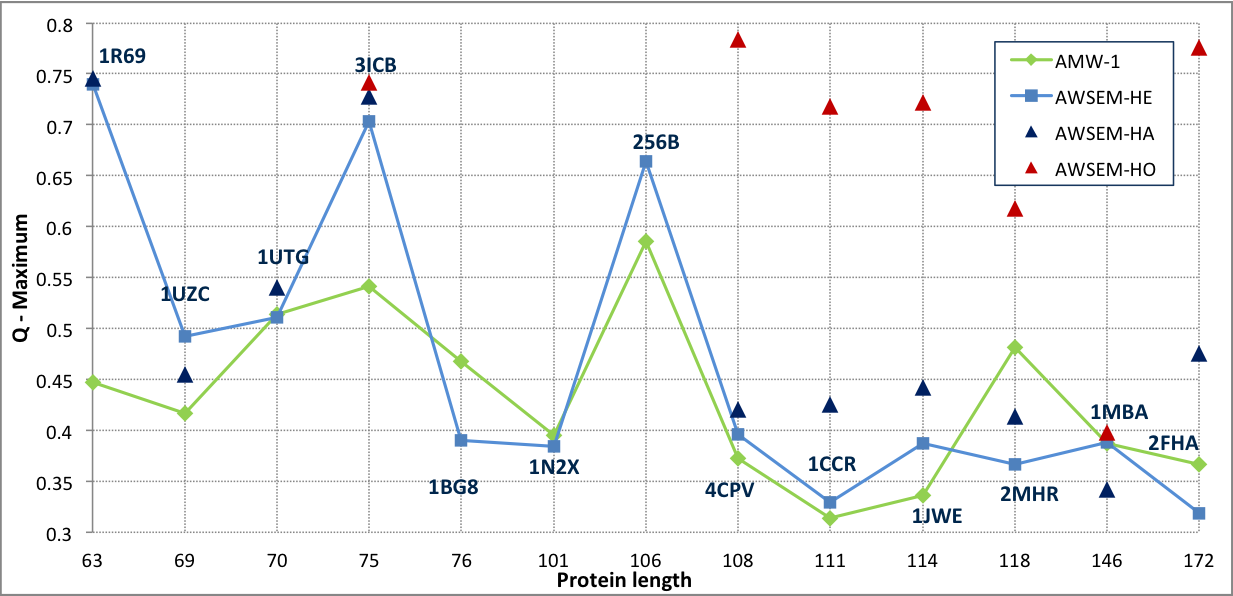}}
\caption{\label{fig:monomerpredictionsummary} The maximum $Q$ sampled during annealing for each target is plotted against sequence length for three different choices of allowed degree of homology between the target and associative memory database structures. ``Homologs excluded" is shown in light blue squares, ``homologs allowed" is shown in dark blue triangles and ``homologs only" is shown in red triangles. For comparison, a previous set of results from \cite{hegler2009restriction} is shown in green diamonds. This figure was adapted from \cite{davtyan2012awsem}.}
\end{center}
\end{figure}

How well can the structures of monomeric proteins be predicted today? We tested the ability of AWSEM to predict the structures of single domain proteins when varying degrees of homology information was assumed to be known. The associative memory interactions come from the structures in the database of ``memories'' to bias local in sequence structures and thus exercises assuming no homology ({\it ab initio}) or acknowledging homology can be set up~\cite{davtyan2012awsem}. The actual native structure of target sequence was never used to inform any of the predictions we discuss below. Figure~\ref{fig:monomerpredictionsummary} shows a summary of the results.

\begin{figure}[h]
\begin{center}
{\includegraphics[width=0.4\textwidth]{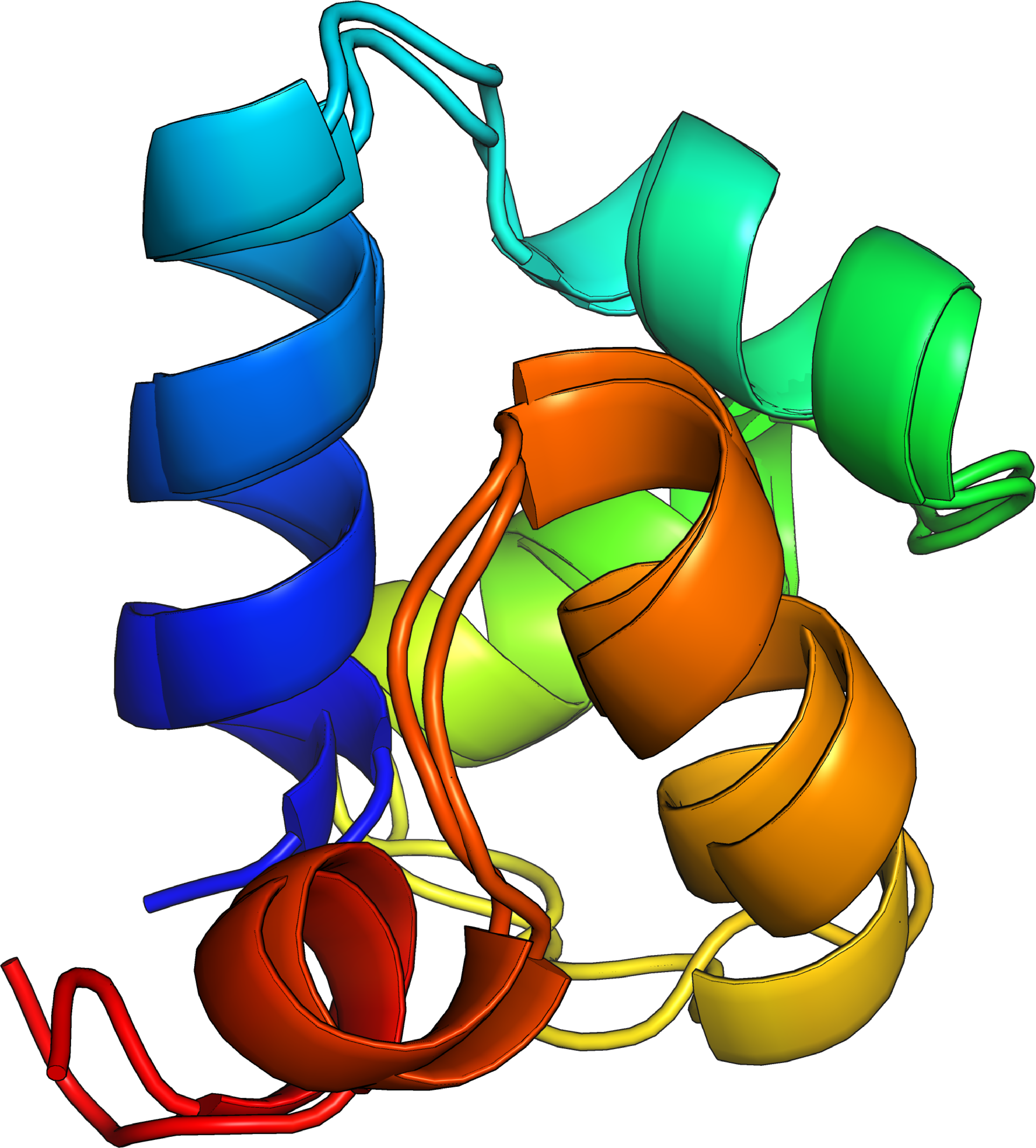}}
{\includegraphics[width=0.4\textwidth]{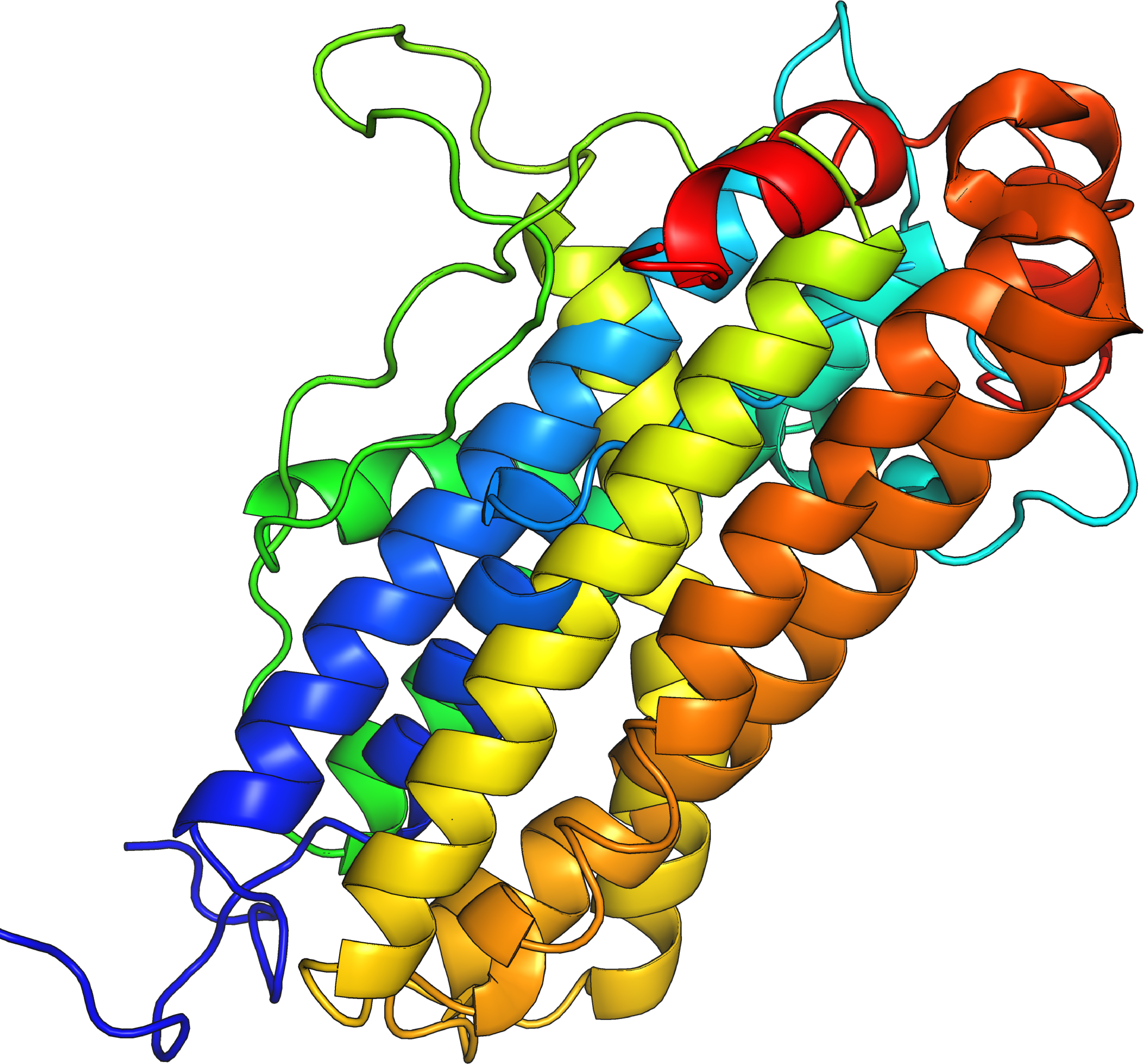}}
\caption{\label{fig:predictedstructures} The highest sampled $Q$ structures for 1r69 (left) and 2fha (right) using the AWSEM model with a homologs excluded associative memory database are shown aligned to their native structures. For the smaller of the two proteins, the predicted and native structures have nearly identical backbone structures. The larger protein has significant similarity to the native structure but shows defects in some of the local structure as well as the packing of helices. This figure was adapted from~\cite{davtyan2012awsem}.}
\end{center}
\end{figure}

The results in Figure~\ref{fig:monomerpredictionsummary} were obtained by performing multiple independent simulated annealing runs starting from an unfolded and extended conformation. The ``homologs excluded'' results correspond to {\it de novo} predictions in the sense that no structural information from any homologous sequence was used to inform the prediction. Homologs are rigorously excluded from the memory list in this exercise. A few example structures predicted without any homology input are shown aligned to their native structures in Figure~\ref{fig:predictedstructures} for the smallest (1r69) and largest (2fha) proteins studied. These structures are close to what a simple matching with a homolog would give if it were to be made available. As expected, including more homology information (``homologs allowed'') improves the structure prediction quality but only modestly if these are not recognized as homologs to start with. If homologs are already recognized, we can then employ ``homologs only'' as memories. For sequences where homologous sequences had experimentally solved structures, using exclusively structures from homologous sequences in the database of associative memories to bias the local structure formation significantly improves the best sampled structures. The best sampled structures when homologs only are used as short range input were always within a few $\AA\ $ of the best structures produced by MODELLER~\cite{fiser2003modeller}, a popular all-atom homology modeling tool.

\clearpage

\subsection{Natural versus Designed Protein Landscapes, Full versus Simplified Amino Acid Alphabets}

The AWSEM potential takes amino acid identity seriously but we have seen the interaction matrix is not maximally complex ({\it i.e.}, only 10 principal components can be used to largely reconstitute the interaction matrix).  Can a simplified folding alphabet then be used, as in the earlier concept of ``biological symmetries''?  Likewise, is evolution the main story (through the fragments) or can proteins designed by humans also be predicted?  To address these issues, the energy landscapes of evolved and some designed proteins were studied using AWSEM.  To look at specific kinetic issues we also employed a non-additive structure based model that has a higher (and more realistic) degree of cooperativity than AWSEM now has~\cite{truong2013funneling}.  The designed sequences chosen for the study were Top7, from the Baker group, and two sequences designed and synthesized by the Takada group~\cite{kuhlman2003design,jin2003novo}.  Top7 was designed to fold to a novel topology starting from a ``sketch'' of the topology and its initial sequence was generated by using fragments with consistent secondary structure~\cite{kuhlman2003design}.  The design procedure was then iterated by the Baker group using Monte Carlo based sequence design and gradient based backbone optimization for multiple rounds.  The two sequences designed by the Takada group began with a target scaffold of a relaxed structure of protein G-related albumin binding domain and then sequences that were expected to fold to this structure were found by a search in sequence space motivated by two criteria inspired by landscape theory~\cite{jin2003novo}.  One sequence, which we call TakadaE, was designed based on a scheme that merely minimized the target structure energy over sequences, while the other sequence, TakadaZ, was designed using a $T_f/T_g$ criterion.  Takada's designs were based on an energy function for structure prediction that used many of the landscape tools and ideas we have already sketched.  Analysis of Top7 in the study was carried out alongside S6, a natural protein having similar secondary structure elements but a different wiring.  Top7 has been the focus of other theoretical investigations that reach similar conclusions to our own study~\cite{zhang2009native, zhang2010competition}.  We also then compared the designs TakadaE and TakadaZ to the behavior of the natural sequence of protein G-related albumin binding domain, which we refer to as TakadaN.  We also can study with AWSEM how robust the of folding of these sequences is to simplification of their amino acid code.

\begin{figure}[h]
\begin{center}
{\includegraphics[width=0.8\textwidth]{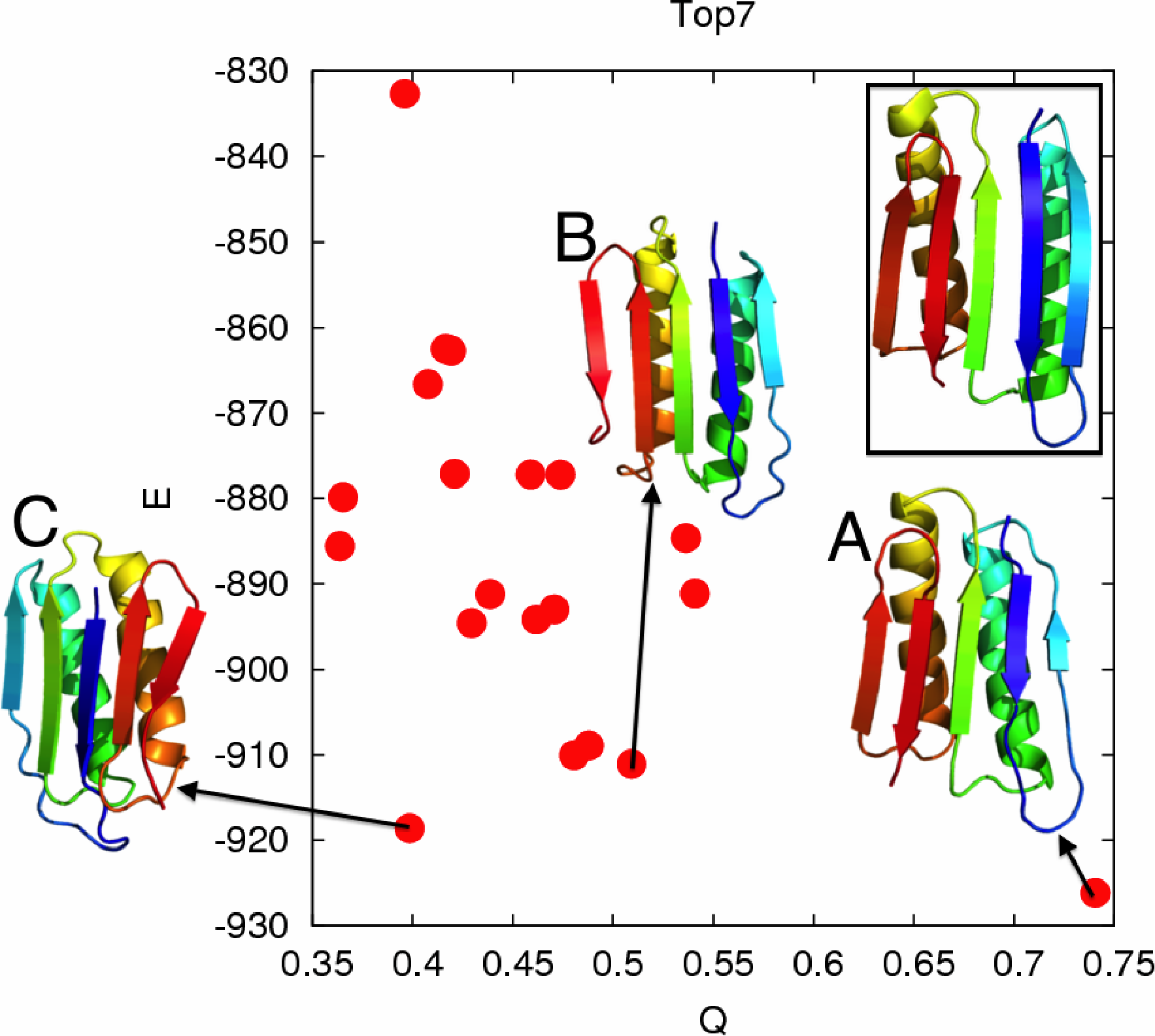}}
\caption{\label{fig:top7alternatestructures} Several energetically competitive non-native structures were found via simulated annealing of the AWSEM ``homologs excluded'' model for Top7. These misfolded states correspond to $\beta$-strand mispairings and are consistent with the notion that the kinetics may be complicated by transitions between compact states. This figure was adapted from~\cite{truong2013funneling}.}
\end{center}
\end{figure}

AWSEM with memories derived from bioinformatic alignments (while excluding fragments from homologous sequences, ``homologs excluded'') was applied to predict the structure of Top7 and S6 via simulated annealing.  For Top7, we found several predicted structures which were of excellent quality, the best structure having 2.1$\AA\ C_{\alpha}$ RMSD (Figure~\ref{fig:top7alternatestructures}). To study kinetics, a flavor of AWSEM which uses the native structure as the only ``memory'' for the short range in sequence associative memory interaction was used to compute free energy profiles as functions of radius of gyration ($R_g$) and fraction of native pairwise distances, $Q_w$ of Top7 and S6.  The free energy profiles looked remarkably similar for Top7 and S6 with the only distinguishing feature being a modestly larger range of $R_g$ over which Top7 is low in free energy.
We also found, however, several topologically distinct structures which were comparable in energy to the best predicted structure.  These structures, characterized by mispairing of $\beta$ strands, are consistent with the suggestion of Baker's group from their kinetic studies~\cite{watters2007highly} and the discussion of their possible role by Chan~\cite{zhang2009native,zhang2010competition}.  The quality of S6 structure prediction is low in comparison to that observed in the survey discussed in the previous section.  We concluded that this lower quality was primarily attributed to a poor bioinformatic prediction of secondary structure, an input to the model which biases AWSEM's Ramachandran potential and $\beta$ hydrogen bonding terms.  AWSEM was able to successfully predict the structures of TakadaE, TakadaZ, and TakadaN.  The quality of the predictions for natural sequence was the highest.  The better structure prediction quality of TakadaE when compared to TakadaZ was largely due to greater funneling of the contact energy for TakadaE.  Apparently the AWSEM energy function is not completely correlated with the one used in the original design by Takada.

\begin{figure}[h]
\begin{center}
{\includegraphics[width=0.8\textwidth]{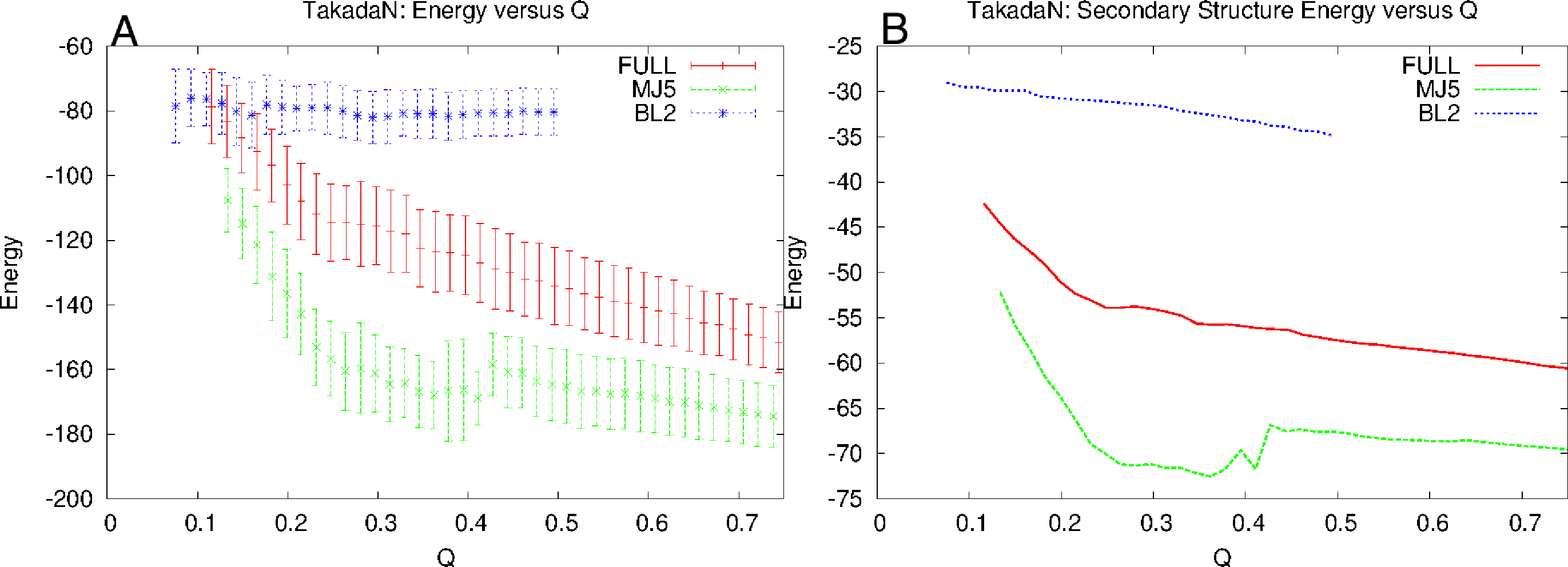}}
\caption{\label{fig:takadaNsimplifiedenergy} This figure shows energy landscape analysis on three versions of the TakadaN protein: full twenty amino acid type sequence (red), MJ5 five letter reduced sequence (green), and BL2 two letter reduced sequence (blue). Both the total energy (left) and secondary structure energy (right) in AWSEM is funneled to high $Q$ for the full sequence. The five letter sequence has an energetic trap at low $Q$, which was found to come from a competing secondary structure. The two letter sequence has a very rugged and flat landscape, and is therefore practically unfoldable. This figure was adapted from~\cite{truong2013funneling}.}
\end{center}
\end{figure}

Modern natural proteins use twenty types of amino acids, but many evolved proteins must, in addition to being foldable, be functional and bind to other proteins, so perhaps that is why the folding palette is so complex.  The evolutionary constraints on function are possible sources of energetic frustration in folding~\cite{ferreiro2007localizing}.  Simplification schemes to two letter and to five letter codes were adapted from Wang {\it et al.}~\cite{wang1999computational,li2003reduction} and the structures of the simplified sequences were predicted with AWSEM.  These structure prediction studies suggest that a two-letter code would be insufficient to fold either the natural or the designed proteins, but a five-letter code may indeed be sufficient for designed sequences.  The structure prediction quality for the five letter simplified variant of TakadaN was considerably poorer than the native sequence, whereas the predictions for five letter simplified variants of Top7, TakadaE, and TakadaZ do not degrade in quality with respect to that found for their respective full sequences.  TakadaN's poorer structure prediction quality can be explained by looking at the total energy as a function of $Q_w$ (Figure~\ref{fig:takadaNsimplifiedenergy}).  While the full sequence is funneled to high Q, the five letter sequence displays an energetic trap, which arises from competing secondary structures. 

\clearpage

\subsection{Prediction of Protein Binding Sites and Structure}
\begin{figure}[h]
\begin{center}
{\includegraphics[width=0.8\textwidth]{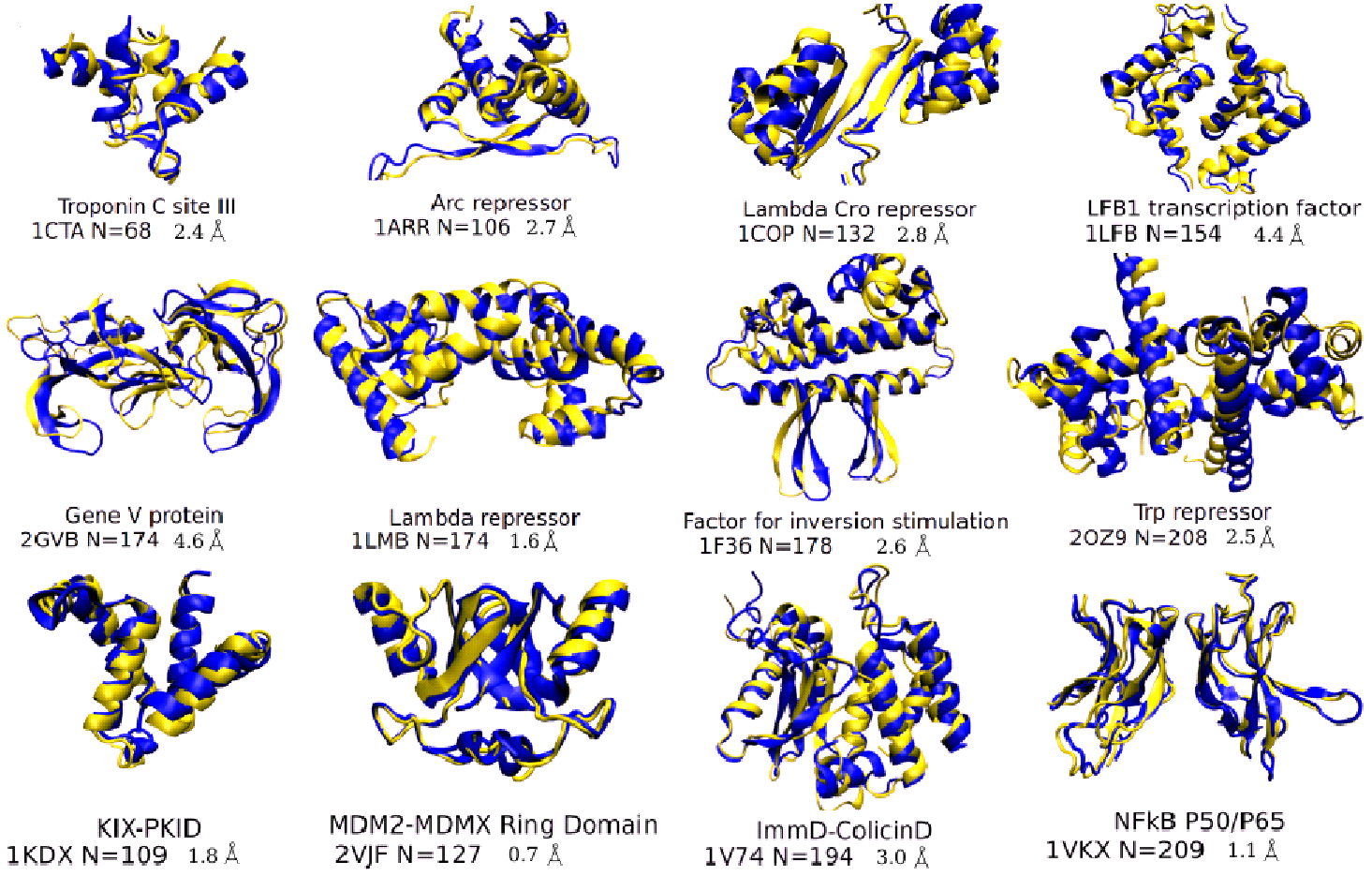}}
\caption{\label{fig:dimer_pred}Snapshots of best predicted structures
(yellow) using AWSEM, compared with the PDB structure (blue). The
name of the proteins, the PDB ID, the number of residues, and the RMSD for the
$C_{\alpha}$ atoms of the predicted complex compared with the PDB structure, are
shown. The first 8 dimers are homodimers, the last 4 are 
heterodimers. This figure was adapted from~\cite{zheng2012predictive}.}
\end{center}
\end{figure}

The water mediated interaction was introduced by the landscape 
analysis of protein binding interfaces by Papoian et al.~\cite{papoian2003role}.
Recently we have investigated the dynamics of protein-protein association using
the Associative-memory, Water mediated, Structure and Energy Model
(AWSEM), a coarse-grained protein folding model which tests those ideas
inspired by the Principle of Minimal Frustration.
The parameters used in the AWSEM code were optimized by maximizing the ratio of the folding temperature to 
the glass transition temperature in order to create funneled folding landscapes for individual monomeric proteins,
even though that optimization started with information about shuffled interfaces in dimers.
Can dimer interfaces still be predicted?
As shown in Figs.~\ref{fig:dimer_pred}, simulated annealing using the AWSEM code is indeed able to predict successfully
the native interfaces of 8 homodimers and 4 heterodimers; thus, AWSEM amounts to a flexible
docking algorithm~\cite{zheng2012predictive}. In these examples, the memory terms use local structural information about the monomers,
much like the ``homologs only'' example discussed above.
The success of the model in predicting binding sites and complete binding structures 
while the training set contains only monomers 
buttresses the idea that the same energy
landscape principles that are applicable monomeric folding also apply to binding processes.

\begin{figure}[h]
\begin{center}
{\includegraphics[width=0.8\textwidth]{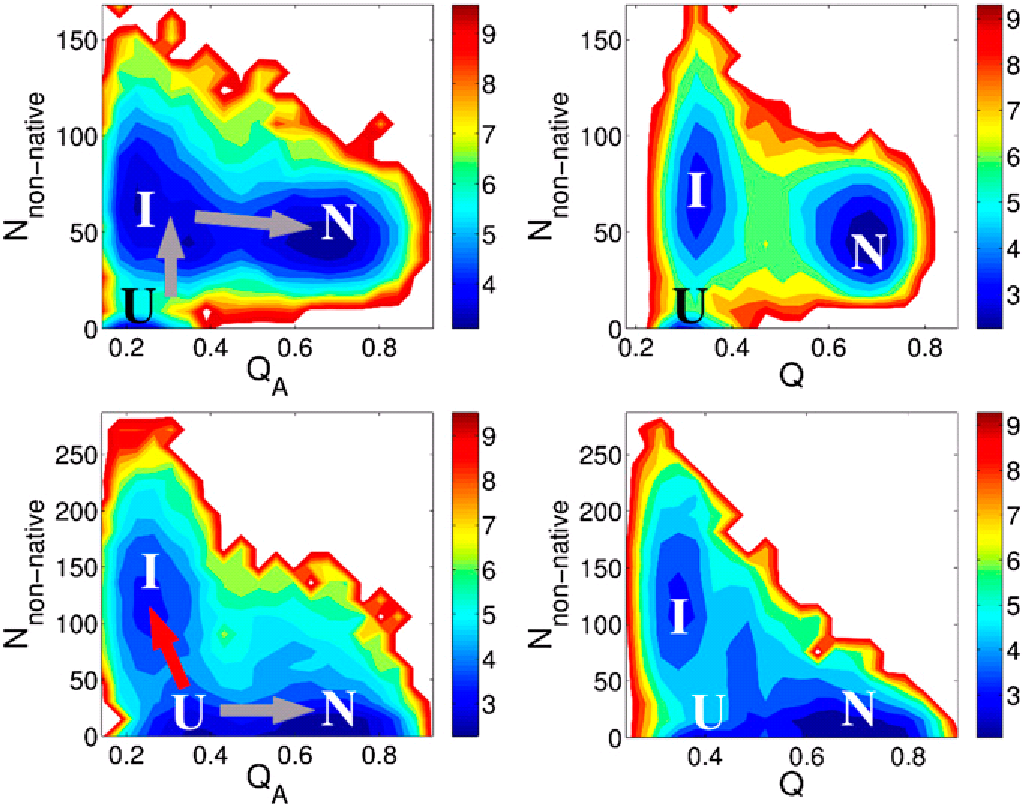}}
\caption{\label{fig:dimer_pmf2d} For Arc repressor (top) and Lambda repressor (bottom),
free energy surfaces at the folding temperature are plotted
as a function of the number of non-native intermonomeric contacts $N_{non-native}$,
 $Q_{A}$ and $Q$ of the complex. $I$, $U$ and $N$ stand for
intermediate, unbound and native bound states, respectively. This figure was adapted from~\cite{zheng2012predictive}.}
\end{center}
\end{figure}

In addition to allowing interface and binding site prediction, 
the potential also allows us 
to study the role of non-native intermonomeric contacts in the process of dimer formation. 
Homodimers are often categorized as being either
obligatory or nonobligatory dimers, meaning that the monomers must
associate in order to complete their folding (obligatory) or are stably folded even in
isolation at physiological temperature (nonobligatory).
Non-native interactions play different roles for
obligatory and non-obligatory dimers as seen in Fig.~\ref{fig:dimer_pmf2d}.
An example of the free energy profile of an obligatory dimer, Arc repressor, is shown on
the top of Fig.~\ref{fig:dimer_pmf2d}. States stabilized by non-native
interactions correspond to on-pathway intermediates that catalyze
the association process through a fly-casting mechanism~\cite{shoemaker2000speeding}; 
the individual monomers, which are both in extended conformations before the association,
have significantly larger capture radii than those of the folded monomers.
The large capture radius increases the rate of binding. In the
case of non-obligatory dimers as in the lower panel of Fig.~\ref{fig:dimer_pmf2d}, however, the states with non-native
contacts generally appear to be off-pathway and impede binding by acting as kinetic traps.

\clearpage

\subsection{Misfolding and frustration}
As we have seen in optimizing potentials for structure prediction,
protein folding and misfolding bear a yin-yang
relationship in the energy landscape theory approach~\cite{wolynes2005energy}.
Can we say anything using predictive landscapes about misfolding in the laboratory?
From a purely physical viewpoint, 
the driving forces must essentially be of the same type for misfolding as those for proper folding.
For evolved proteins, which satisfy the Principle
of Minimal Frustration, domain-swapped interactions are therefore the most obvious
candidate for specific interactions that drive misfolding~\cite{yang2004domain}. 
Like their counterparts in the monomer, the native contacts, domain-swapped contacts
are in general stronger than other contacts. 
These same strong interactions can also
drive oligomerization via domain-swapping with nearby domains, as suggested in~\cite{bennett2006deposition}.
Misfolding can occur, however, in a different way. AWSEM simulations show that the formation of self-recognition contacts,
which are strong contacts formed between the same sequence segments from different
polypeptides~\cite{sawaya2007atomic}, is also a possibility.
These self-recognition contacts between two segments
in different molecules can be extremely strong since the segments are pretty rigid locally
and these strong interactions can act like Velcro to hold two different molecules together.
Unlike domain swapping interactions, self-recognition contacts 
have no exact counterpart in the native structure
and therefore can only be involved in misfolding.
A survey shows they have been avoided but not completed eliminated by evolution~\cite{schwartz2001frequencies,goldschmidt2010identifying}.

\begin{figure}[h]
\begin{center}
\includegraphics[width=0.8\textwidth]{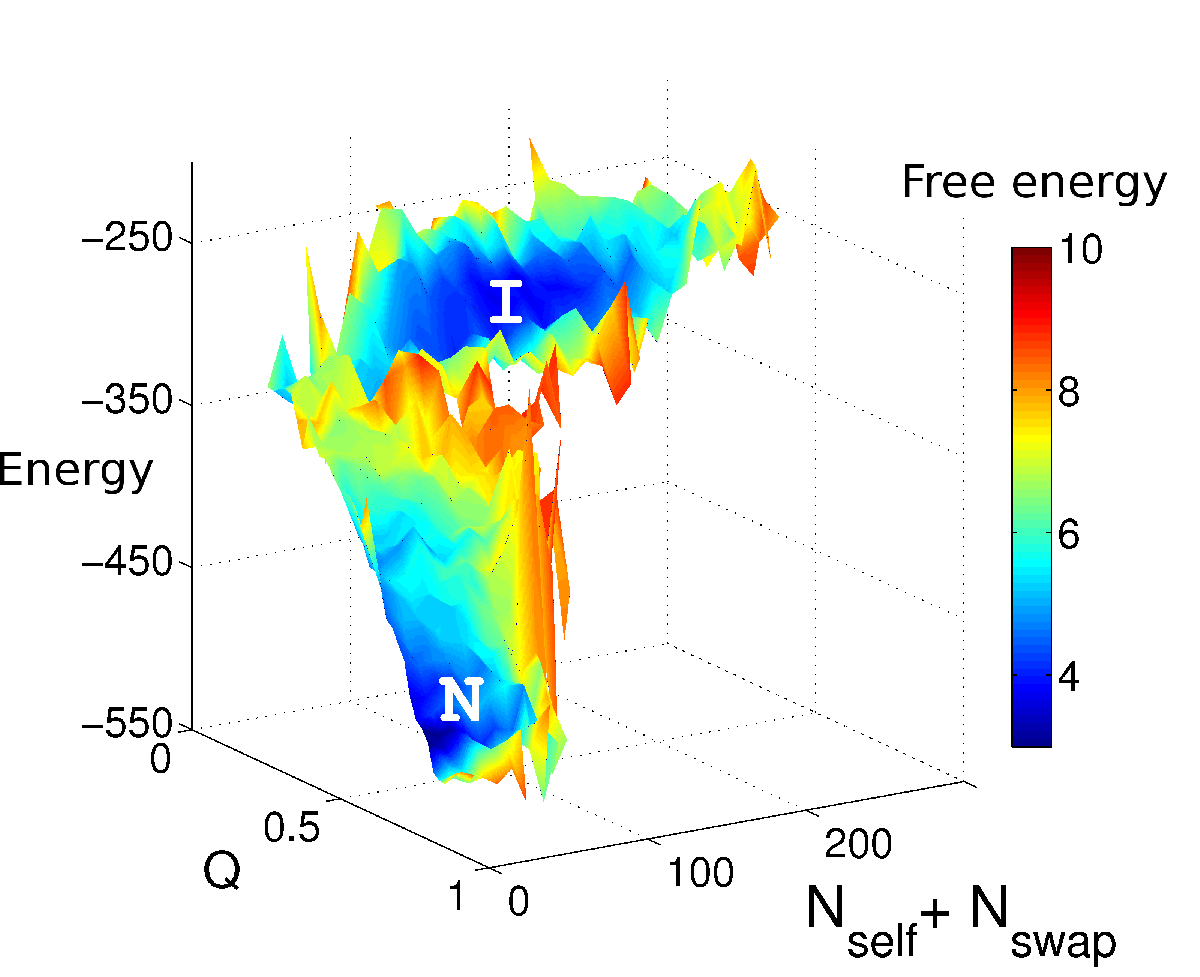}
\caption{\label{fig:misfold_pmf4d}Energy and free energy surfaces for I27-I27 at
its folding temperature. $N_{self}$ and $N_{swap}$ are the number
of self-recognition contacts and the number of domain-swapped contacts,
respectively. The trapped states $I$ have higher energies than the
native states $N$, as shown in the z-axis, but have similar free
energies as the native states, as shown by the color coding of the
free energy, with scale indicated in the side bar. We see that the
ensemble $I$ states are entropically favored. As temperature increases,
the intermediate ensemble will become more stable than the native 
ensemble. This figure was adapted from~\cite{zheng2013frustration}.}
\end{center}
\end{figure}

These ideas were explored using the AWSEM code by studying
fused dimers consisting of the 27th Ig domain of human cardiac titin (I27; PDB ID 1TIU). 
Fig.~\ref{fig:misfold_pmf4d} shows the energy and free energy profiles for 
the I27-I27 fused dimer using two order parameters: 
the fraction of native contacts $Q$ and the sum of the number of self-recognition contacts
$N_{self}$ and the number of swapped contacts $N_{swap}$. The misfolded
state I is energetically less stable than the native state N but is entropically
more favored since it is disordered in the other non-Velcro parts of the molecule.
This type of metastable ensemble acts as a
kinetic trap even below the folding temperature. When a fused dimer
construct becomes trapped in this metastable state, it can act to initiate 
aggregation when other fused dimers are present nearby.

\clearpage

\subsection{Initiation and branching in aggregation}
\begin{figure}[h]
\begin{center}
\includegraphics[width=0.8\textwidth]{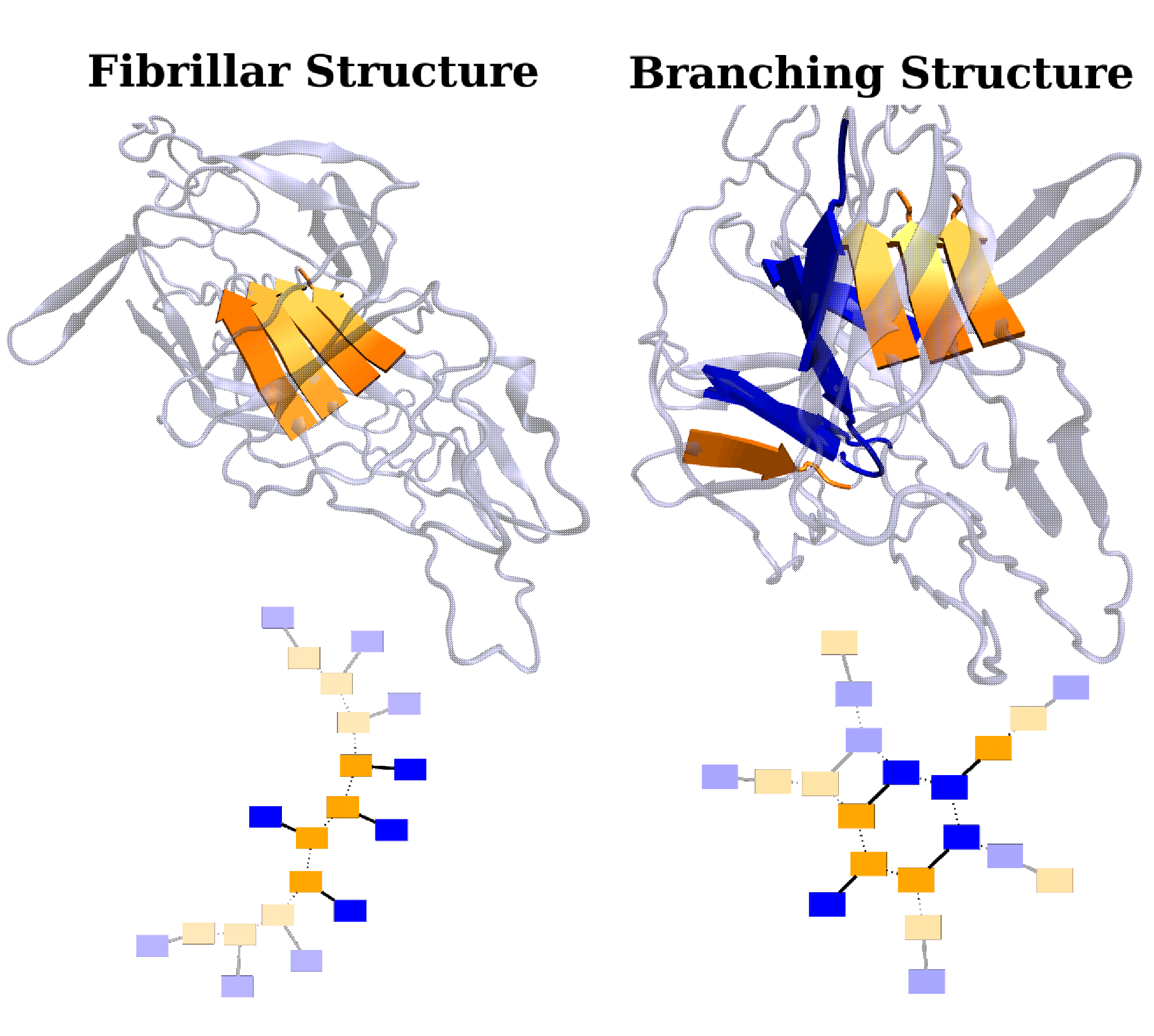}
\caption{\label{fig:Misfold-tetramer}Two types of misfolded structures of the wild-type tetramer are shown in three-dimensional (top) and simplified two-dimensional (bottom) representations. In the 2-D model, bold colors indicate the actual structures found in the AWSEM molecular dynamics simulations and the light colors are examples of how these structures might further develop in the presence of more protein copies. In each protein, there are two sticky segments, shown in orange and blue. A solid line represents the rest of each protein. Dashed lines represent stabilizing interactions formed between two sticky segments from different proteins. A fibrillar structure is shown on the left and a branching structure is shown on the right. The presence of two or more sticky segments in one protein allows for a greater diversity of possible misfolded structures. This figure was adapted from~\cite{Zheng27112013}.}
\end{center}
\end{figure}

Mature fibrils are a very striking feature found in patients suffering from protein
aggregation related diseases, but how these structures relate to disease
pathology remains open~\cite{ross2004protein,chiti2006protein,benilova2012toxic,berthelot2012does}.
Recent evidence supports the notion that in some cases fibers are
byproducts of a process that starts with oligomers that are themselves toxic and
pathogenic. It is then vitally important to understand the early stages
of aggregation that are invisible to many experimental techniques.
AWSEM simulations are proving useful for proposing candidate misfolded
structures and other oligomeric species. Recently we have studied this problem for higher oligomers
of I27. In the sequence of I27, there are two segments that can recognize themselves.
According to the AWSEM energy function~\cite{zheng2013frustration}, one
of these is weaker than the other. AWSEM simulations show that misfolded
multidomain structures can arise from multiple independent polymer
units that are cross-linked to other chains via these self-recognition
interactions. Because $\beta$-strands can be linked to two other
$\beta$-strands via backbone hydrogen bonding, a single domain of I27 can
be thought of as a polymer unit having two reactive groups~\cite{flory1942constitution,stockmayer1943theory}
corresponding to the self-recognizing fragment.
In the case where there is only a single self-recognition segment,
each chain can only form two crosslinks. Therefore the polymerization
would proceed linearly through a combination of elongation and breakup
events, and later through the association of protofibril structures
into mature fibrils. The simulations show~\cite{Zheng27112013}, however, that when the number of reactive groups per
chain exceeds 2, as is the case here, branched structures become possible
and completely ordered fibrils must compete with these other structures.
In AWSEM simulations, simulated annealing of the tetramer yielded protofibril-like 
structures 30\% of the time and branched structures 33\% of the time. 
The fibrillar and branched structures are illustrated in Fig.~\ref{fig:Misfold-tetramer}.
The growth kinetics and molecular weight distribution of branched and linear aggregates
should be very different, and this difference should be discernible
in experiment. Structure prediction algorithms thus can give inspiration to new
studies of the misfolding kinetics. The specificity of the Velcro self-recognizing
segments and the local structural uniqueness of the oligomers also give us hope
for druggability of these misfolded structures with other peptides~\cite{sievers2011structure,yang2013transthyretin} or small molecules.

\clearpage

\section{Conclusions and outlook}
\label{sec:conclusionsandoutlook}

The fundamental questions regarding how single domain proteins fold have been answered by the Principle of Minimal Frustration in the framework of statistical energy landscape theory. The fidelity of atomistic models of proteins is now good enough that small proteins can be folded~\cite{lindorff2011fast}, and careful analysis shows the results of these folding simulations are in quantitative agreement with what is expected on the basis of the Principle of Minimal Frustration~\cite{best2013native}, but the computational complexity of these full atomistic models makes simulations of biologically interesting phenomena far from routine. Fortunately, the quantitative formulation of the principle of minimal frustration, and the resulting algorithms that use this quantitative formulation to parameterize coarse-grained models have been successfully developed and evaluated. These models can predict the structure of modest size monomeric proteins as well as protein-protein binding sites and dimeric interfaces and bound structures. Having passed the structure prediction test, these models give insights into questions regarding the aberrant kinetics of designed proteins. They can also be used to elucidate the molecular origin of misfolding in multidomain proteins and formulate new hypotheses about the nature of oligomer structures that initiate aggregation. There are very many possible applications for optimized coarse-grained models, spanning almost the complete range of topics in molecular biophysics. Evolution has already gone through the trouble of correcting its ``first 5000 mistakes'', and energy functions have likewise gone through a rigorous selection process, resulting in models that capture the essential physics without over complicating the computations. Further application and development of these models promises to deliver a wealth of information regarding the nature of life at its most basic level.

\section{Acknowledgments}
\label{sec:acknowledgements}
This paper reviews more than twenty five years of work on structure prediction and folding theory in the Wolynes group that has been supported by the NSF and the NIH. In recent years, the funding was from NIH R01 GM44557 and NIH P01 GM071862. The computational support, especially from the Center for Theoretical Biological Physics and earlier the University of Illinois NCSA, is gratefully acknowledged. Many essential co-workers in these endeavors are acknowledged in the references, but PGW would like to especially single out Zan Luthey-Schulten for her long-standing contributions to the effort. Support from the Bullard-Welch Chair at Rice University has also been essential in recent years.

\newpage

\section{Appendix}
We wish to maximize $\frac{T_f}{T_g}$ or equivalently, $\frac{\delta E_s}{\Delta E}$.  Optimization is simplest when considering parameters in the energy function that enter in a linear fashion, $E = \sum_i{\gamma_i\phi_i}$.  The $\gamma_i$'s are the strengths of the interaction terms whereas the $\phi_i$'s are the basic forms of the interaction potential.  The stability gap can be written as $\delta E_s=\bm{A\gamma}$ whereas the energetic variance can be written as $\Delta E^2=\bm{\gamma B \gamma}$.  $\bm{A}$ and $\bm{\gamma}$ are vectors of dimensionality equal to the number of interaction types while $\bm{B}$ is a matrix.  $\bm{A}$ and $\bm{B}$ are defined as \\
\begin{align*}
A_i&=\left<\phi_i\right>_{mg}-\phi_n\\
B_{i,j}&=\left<\phi_i\phi_j\right>_{mg}-\left<\phi_i\right>_{mg}\left<\phi_j\right>_{mg}
\end{align*}
Optimization of $\frac{\delta E_s}{\Delta E}=\bm{\frac{A\gamma}{\sqrt{\gamma B \gamma}}}$ with respect to $\gamma_i$ is equivalent to maximization of $\bm{A\gamma}$ under the linear constraint that $\bm{\sqrt{\gamma B \gamma}}$ is a constant.  Starting from the condition\\
\\
\begin{align*}
\frac{\partial F}{\partial\bm{\gamma}} = \frac{\partial}{\partial\bm{\gamma}}\left[\bm{A\gamma}-\bm{\sqrt{\gamma B \gamma}}\right] = 0
\intertext{we find for an individual $\gamma_k$ the following equations:}
\end{align*}
\begin{align*}
\frac{\partial}{\partial\gamma_k}\left[\sum_iA_i\gamma_i-\mu\sqrt{\sum_{i,j}\gamma_iB_{ij}\gamma_j}\right]&=0\\
A_k-\mu\left[\frac{1}{2}\frac{1}{\sqrt{\sum_{i,j}\gamma_iB_{ij}\gamma_j}}\frac{\partial}{\partial\gamma_k}\sum_{ij}\gamma_iB_{i,j}\gamma_j\right]&=0\\
\intertext{The quadratic form is easily differentiated to yield}
\end{align*}
\vspace{-2.3cm}
\begin{align*}
\frac{\partial}{\partial\gamma_k}\sum_{i,j}\gamma_iB_{i,j}\gamma_j&=
\frac{\partial}{\partial\gamma_k}
\begin{bmatrix}
\gamma_1B_{1,1}\gamma_1+...+\gamma_1B_{1,k}\gamma_k+..+\gamma_1B_{1,n}\gamma_n\\
\gamma_kB_{k,1}\gamma_1+...+\gamma_kB_{k,k}\gamma_k+..+\gamma_kB_{k,n}\gamma_n\\
\gamma_nB_{n,1}\gamma_1+...+\gamma_nB_{n,k}\gamma_k+..+\gamma_NB_{n,n}\gamma_n
\end{bmatrix}\\
&=2\gamma_kB_{k,k}+\sum_{i\neq k}\gamma_iB_{i,k}+\sum_{j\neq k}B_{k,j}\gamma_j\text{ but $B$ is symmetric}\\
&=2\sum_jB_{j,k}\gamma_j
\intertext{The term $\frac{1}{\sqrt{\sum_{i,j}\gamma_iB_{ij}\gamma_j}}$ is a scalar which can be absorbed into the new Lagrange multiplier, $\mu^*$ and thus we find}
\end{align*}
\vspace{-2.3cm}
\begin{align*}
A_k-\mu^*\sum_jB_{j,k}\gamma_j=0
\intertext{which can be generalized to all the other components}
\end{align*}
\vspace{-2.3cm}
\begin{align*}
\begin{pmatrix}
  A_{1,1} \\
  A_{2,1} \\
  \vdots \\
  A_{n,1}
 \end{pmatrix}-2\mu
\begin{pmatrix}
  B_{1,1} & B_{1,2} & \cdots & B_{1,n} \\
  B_{2,1} & B_{2,2} & \cdots & B_{2,n} \\
  \vdots  & \vdots  & \ddots & \vdots  \\
  B_{n,1} & B_{n,2} & \cdots & B_{n,n}
 \end{pmatrix}
\begin{pmatrix}
  \gamma_{1,1} \\
  \gamma_{2,1} \\
  \vdots \\
  \gamma_{n,1}
 \end{pmatrix}=0
\end{align*}
\begin{align*}
\intertext{and finally in matrix notation, one obtains}
\bm{A}-\mu^*\bm{B\gamma}=0\\
\bm{\gamma}=\frac{1}{\mu^*}\bm{B^{-1}A}
\end{align*}

\newpage

\bibliographystyle{abbrv}
\bibliography{learningtofoldproteins}
%\nocite{*}

\end{document}